\documentclass{aa}

\usepackage{graphicx}
\usepackage{txfonts}

\usepackage{wasysym}

\usepackage[percent]{overpic}

\usepackage[pdftitle={Exomoon indicators in high-precision transit light curves}, colorlinks = true, breaklinks = true, citecolor = blue, linkcolor = blue, urlcolor = blue, pdfauthor = {Kai Rodenbeck}]{hyperref}
\usepackage{xcolor}

\usepackage{rotating}

\graphicspath{{./figs/}}
\begin{document}

\title{Exomoon indicators in high-precision transit light curves}

\author{Kai Rodenbeck \inst{1}
             \and
             Ren\'{e} Heller\inst{2}
             \and
             Laurent Gizon\inst{1,2}
}

\institute{
           Institute for Astrophysics, Georg August University G\"ottingen, Friedrich-Hund-Platz 1, 37077 G\"ottingen, Germany
           \and
           Max Planck Institute for Solar System Research, Justus-von-Liebig-Weg 3, 37077 G\"ottingen, Germany
           \\
           \email{(rodenbeck/heller/gizon)@mps.mpg.de} 
}

   \date{draft: \today}

\abstract
{While the solar system contains about 20 times more moons than planets, no moon has been confirmed around any of the thousands of extrasolar planets known so far. Considering the large computational load required for the statistical vetting of exomoon candidates in a star-planet-moon framework, tools for an uncomplicated identification of the most promising exomoon candidates could be beneficial to streamline follow-up studies.}
{Here we study three exomoon indicators that emerge if well-established planet-only models are fitted to a planet-moon transit light curve: transit timing variations (TTVs), transit duration variations (TDVs), and apparent planetary transit radius variations (TRVs). We re-evaluate under realistic conditions the previously proposed exomoon signatures in the TTV and TDV series.}
{We simulate light curves of a transiting exoplanet with a single moon, taking into account stellar limb darkening, orbital inclinations, planet-moon occultations, and noise from both stellar granulation and instrumental effects. These model light curves are then fitted with a planet-only transit model, pretending there were no moon, and we explore the resulting TTV, TDV, and TRV series for evidence of the moon.}
{The previously described ellipse in the TTV-TDV diagram of an exoplanet with a moon emerges only for high-density moons. Low-density moons, however, distort the sinusoidal shapes of the TTV and the TDV series due to their photometric contribution to the combined planet-moon transit. Sufficiently large moons can nevertheless produce periodic apparent TRVs of their host planets that could be observable. We find that Kepler and PLATO have similar performances in detecting the exomoon-induced TRV effect around simulated bright ($m_V=8$) stars. These stars, however, are rare in the Kepler sample but will be abundant in the PLATO sample. Moreover, PLATO's higher cadence yields a stronger TTV signal. We detect substantial TRVs of the Saturn-sized planet Kepler-856\,b although an exomoon could only ensure Hill stability in a very narrow orbital range.}
{The periodogram of the sequence of transit radius measurements can indicate the presence of a moon. The TTV and TDV series of exoplanets with moons can be more complex than previously assumed. We propose that TRVs could be a more promising means to identify exomoons in large exoplanet surveys.}


\maketitle

\section{Introduction}

Among the many unexpected discoveries of the Kepler mission such as evaporating planets \citep{2012ApJ...752....1R}, exocomets \citep{2018MNRAS.474.1453R}, and even evidence of an exo-Trojan population \citep{2015ApJ...811....1H}, exomoons have hitherto escaped an unambiguous detection. The peculiar case of the exomoon candidate around Kepler-1625\,b \citep{2018AJ....155...36T,2018SciA....4.1784T} is extremely hard to assess based on the very limited amount of data \citep{2018A&A...610A..39H,2018A&A...617A..49R,2019ApJ...877L..15K}.

Many ways have been proposed to look for moons in transit light curves \citep[for a review see][]{2018haex.bookE..35H}. \citet{1999A&AS..134..553S} suggested that a massive moon can distract the transits of its host planet from strict periodicity and induce transit timing variations (TTVs). \citet{2009MNRAS.392..181K} identified an additional effect of the moon on the planet's sky-projected velocity component tangential to the line of sight that results in transit duration variations (TDVs). These effects relate to the planet's position with respect the planet-moon barycenter, which is modeled on a Keplerian orbit around the star, and they depend on the planet-satellite orbital semi-major axis ($a_{\rm ps}$) and on the satellite's mass ($M_{\rm s}$) but they do not depend on the satellite's radius ($R_{\rm s}$). \citet{2016A&A...591A..67H} proposed that the combined planetary TTV and TDV effects can produce distinct ellipsoidal patterns in the TTV-TDV diagram if the moon is sufficiently small to avoid any effects on the photometric transit center.

\citet{2006A&A...450..395S}, however, noted that the stellar dimming caused by a moon with a sufficiently large radius can affect the photometric center of the transit light curve, defined as $\tau = \sum_i t_i {\Delta}m_i / \sum_i {\Delta}m_i $ with $t_i$ as the times when the stellar differential magnitudes (${\Delta}m_i$) are measured. For small (arbitrarily massive) moons, $\tau$ would coincide with the transit midpoint of the planet but for large moons $\tau$ can be substantially offset from the position of the midpoint of the planetary transit. \citet{2007A&A...470..727S} derived an analytical estimate of the photometric TTV effect (dubbed ``TTV$_{\rm p}$'') and compared it to the magnitude of the barycentric TTV effect (dubbed ``TTV$_{\rm b}$''). They define the TTV$_{\rm p}$ as the difference between TTV$_{\rm b}$ and $\tau$.

Here we simulate many different transit light curves of a planet with a moon of non-negligible radius and then fit the resulting data with a planet-only transit model to obtain the TTVs and TDVs of a hypothetical sequence of transits. Our aim is to provide the exoplanet community with a tool to inspect their TTV-TDV distributions for possible exomoon candidates without the need of developing a fully consistent, photodynamical transit modelling of a planet with a moon \citep{2011MNRAS.416..689K,2018A&A...617A..49R}. We show that sufficiently large moons can distort the previously proposed ellipse in the TTV-TDV into very complicated patterns, which are much harder to discriminate from the TTV-TDV distribution of a planet without a moon. We also identify a new effect that appears in the sequence of transit radius ($R_{\rm p}$) measurements, which could be used to identify exomoon candidates in large exoplanet surveys.

\section{Method}

\subsection{Setup of the transit model}

\begin{figure}
    \centering
    \includegraphics{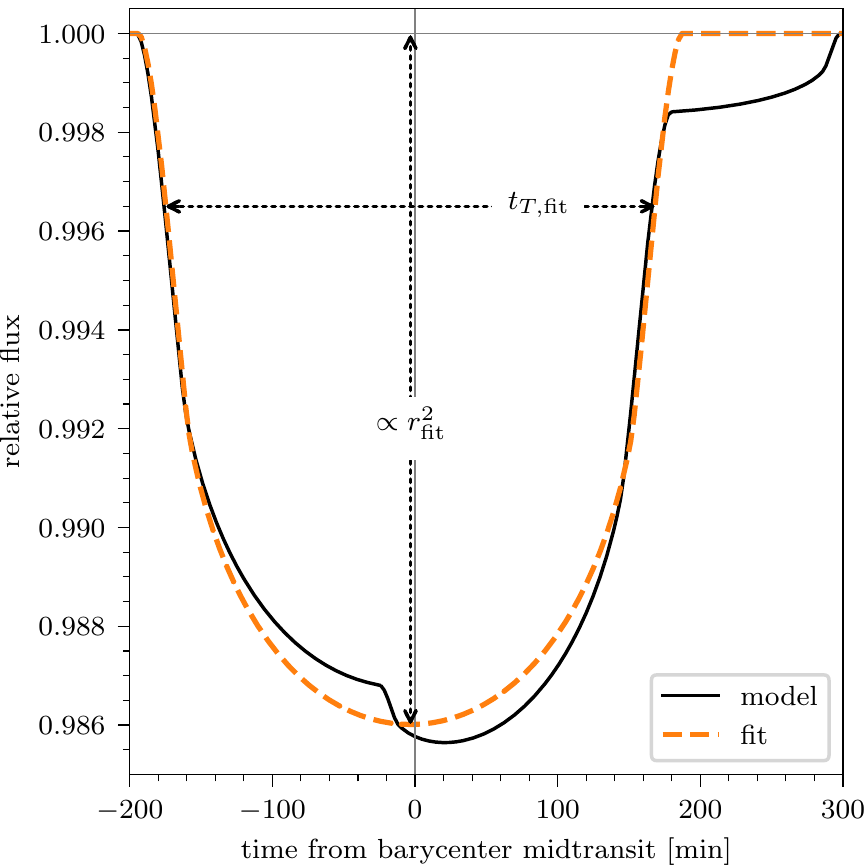}
    \caption{Example of a planet-only fit (dashed orange line) to the planet-moon model (solid black line). The star is sun-like, the planet is Jupiter-sized and in a 30\,d orbit around the star, and the moon is Neptune-sized and in a 3.55\,d orbit around the planet. In this example, the fitted planet-to-star radius ratio ($r_{\rm fit}$), which goes into the calculation of the transit depth with a power of 2, is slightly overestimated due to the moon's photometric signature. As a comparison between the amplitude of the moon signal and the expected noise, the noise from the Poisson statistics of an $m_V$ = 8 star (using a 1\,hr binning) is 12.6\,ppm for PLATO and 6.0\,ppm for Kepler. This is two magnitudes smaller than the amplitude of the moon signature but possibly comparable the effect of stellar noise, e.g. from activity.}
    \label{fig:shape-shift}
\end{figure}

Figure~\ref{fig:shape-shift} demonstrates how the transit light curve of an exoplanet with a moon (solid black line) differs ever so slightly from the light curve expected for a single planet (orange dashed line). For one thing, the midpoint of the planetary transit changes between transits with respect to the planet-moon barycenter, thereby causing a TTV$_{\rm b}$ effect \citep{1999A&AS..134..553S}. Moreover, the additional dimming of the star by a sufficiently large moon could act to shift the photometric center of the combined transit, an effect referred to as TTV$_{\rm p}$ \citep{2006A&A...450..395S,2007A&A...470..727S}. The same principle of the moon's photometric (rather than the barycentric) effect should be applicable to the TDV, though it has not been explored in the literature so far. And finally, Fig.~\ref{fig:shape-shift} illustrates how the planet-only fit to the planet-moon model overestimates the radius of the planet. For single planets, the transit depth is roughly proportional to $R_{\rm p}^2$, although details depend on the stellar limb darkening \citep{2019A&A...623A.137H}. But a moon can affect the radius estimate for the planet and therefore mimic transit radius variations (TRVs) in a sequence of transit measurements.

We investigate both the barycentric and the photometric contributions to the measured TTVs and TDVs of a planet-with-moon transit light curve that is fitted with a planet-only model. To separate the TTV$_{\rm b}$ and TTV$_{\rm p}$ effects, we simulate transits of a hypothetical planet-moon system with a zero-radius or zero-mass moon, respectively, and then fit the simulated light curves with a planet-only transit model. Finally, we assume a realistic moon with both mass and radius and redo our fit to the resulting light curve.

We simulate exoplanet-exomoon transits using our implementation of the \citet{2002ApJ...580L.171M} analytic transit model similar to \citet{2018A&A...617A..49R}. As an improvement to \citet{2018A&A...617A..49R}, however, our model now includes planet-moon occultations and the possibility of inclining the planet-moon orbit with respect to the line-of-sight. The planet-only transit model is fitted to the simulated light curve using the planet-to-star radius ratio ($r_{\rm p}$), transit duration ($t_{\rm T}$), and the orbital phase ($\varphi_{\rm b}$), while the orbital period ($P_{\rm b}$) and the limb darkening coefficients \citep[$q_1$ and $q_2$ for the quadratic limb darkening model as per][]{2013MNRAS.435.2152K} are kept constant at the respective values shown in Table~\ref{tab:params}. Figure~\ref{fig:shape-shift} shows an example of a fitted planet-only light curve (orange dashed line) to the simulated noiseless transit of a planet-moon system.

We also investigate the possibility of detecting TRVs. Variations of the planetary transit depth have been observed before \citep{2016ApJS..225....9H} but to our knowledge they have not been treated in the context of exomoons so far.

\subsection{Simulated noise in the light curves}

We compare the effect of photon noise between two space missions: Kepler \citep{2010Sci...327..977B} and PLATO \citep{2014ExA....38..249R}, the latter of which is currently scheduled for launch in 2026. For both missions we test stars of two different apparent magnitudes, namely 8 and 11. The two stellar magnitudes provide us with two possible amplitudes of the white noise level per data point, which is composed of photon noise and telescope noise. For Kepler long cadence observations with a cadence of 29.4\,min\footnote{Each cadence is composed of a 6.019803\,s exposure and a 0.51895\,s read out time. Long cadence data is composed of 270 such integrations, which yields a cadence of 1765.46\,s or 29.42\,min \citep{Thompson2016}.}, we  obtain a white noise level of 9 parts per million (ppm) at magnitude 8 and 36\,ppm at magnitude 11 \citep{2010ApJ...713L..79K}. For PLATO we estimate 113\,ppm at magnitude 8 and 448\,ppm at magnitude 11, both at a cadence of 25\,s. These estimates for PLATO assume an optimal target coverage by all 24 normal cameras.

We also simulate a time-correlated granulation noise component according to \citet{2011ApJS..197....6G}, where the granulation power spectrum is modeled as a Lorentzian distribution in the frequency domain and then transformed into the time domain. The granulation amplitude and time scale depend on the star's surface gravity and temperature. We use two stellar types, a sun-like star and a red dwarf with similar properties as $\epsilon\,$Eridani ($\epsilon\,$Eri).



%

\subsection{Parameterization of the test cases}
\label{sec:parameterization}

\begin{figure}
    \centering
    \includegraphics[width=0.5\textwidth]{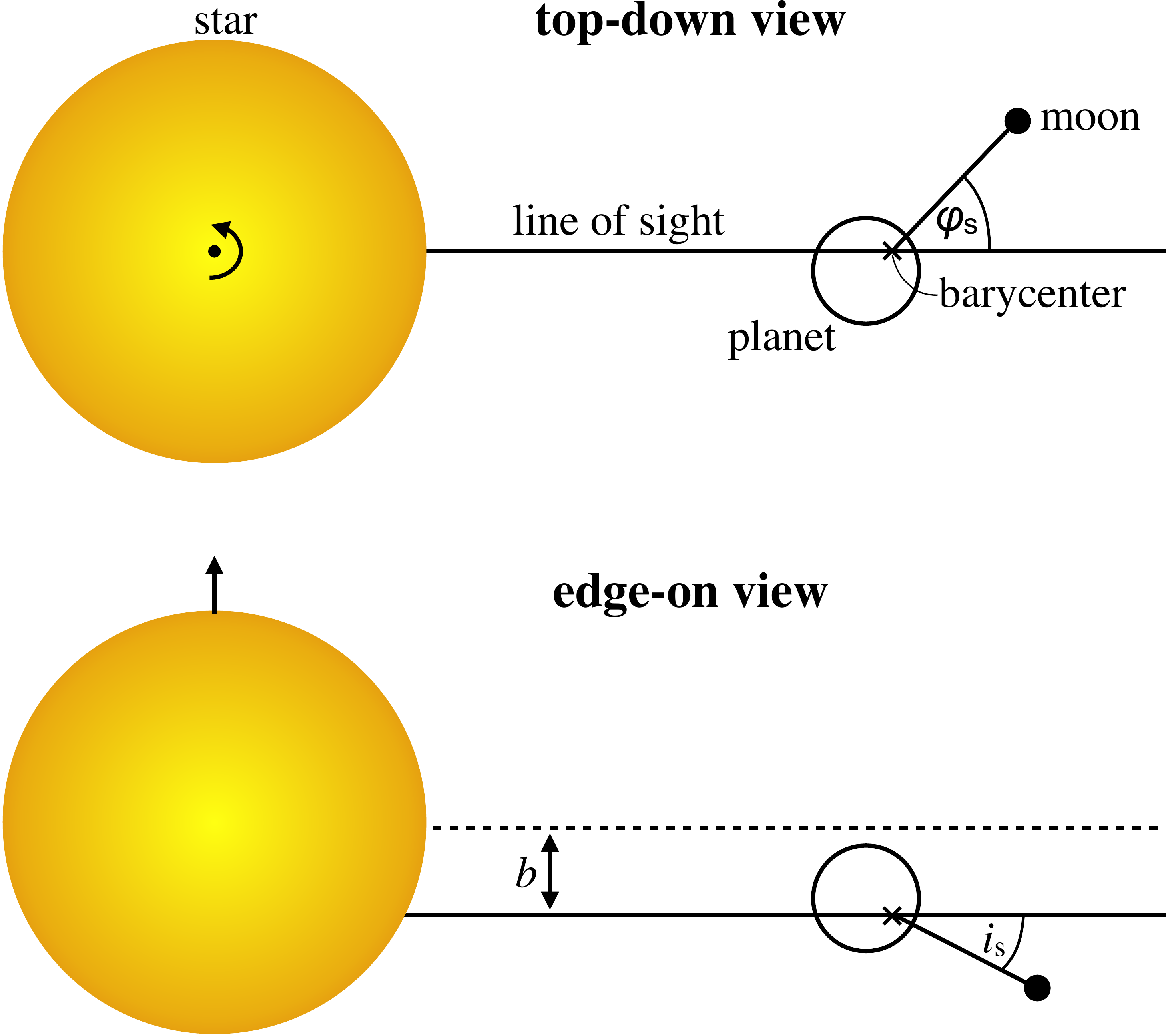}
    \caption{Illustration of the orbital geometry and of our parameterization of the simulated planet-moon systems. $\varphi_{\rm s}$ is the orbital phase relative to the line of sight at the time of mid-transit.}
    \label{fig:diagram}
\end{figure}

We investigate several hypothetical star-planet-moon systems to explore the observability of the resulting TTV, TDV, and TRV effects. We ensure orbital stability of the moons by arranging them sufficiently deep in the gravitational potential of their host planet, that is, at $<0.5$ times the planetary Hill radius. This distance has been shown to guarantee orbital stability of both pro- and retrograde moons based on numerical $N$-body simulations \citep{2006MNRAS.373.1227D}. We choose stellar limb darkening coefficients ($q_1$, $q_2$) for both the sun and a K2V star akin to $\epsilon$\,Eri from \citet{2011A&A...529A..75C}.

In Table~\ref{tab:params} we summarize our test cases. For case 1, and for all of its sub cases 1a - 1p, we choose the planet to be a Jupiter-sized and the moon to be Earth-sized. The orbit period of the moon can take two values, either the orbital period of Europa (3.55\,d) or that of Io (1.77\,d). The barycenter of the planet-moon system is in a $P_{\rm b}=30$\,d orbit around its Sun-like star. In test cases 2a - 2d, we simulate Neptune-sized moons and vary both orbital periods involved to test the effect of the number and duration of the transits. We set $P_{\rm b}$ to be either 30\,d orbit or close to twice that value. In fact, however, we shorten the 60\,d orbit by 1.69\,d to 58.31\,d for the Io-like orbit and by 1.59\,d to 58.41\,d for the Europa-wide orbit in order for the moon to show the same orbital advancement between transits, that is, for the remainder of $P_{\rm b}/P_{\rm s}$ to be the same \citep{2016A&A...591A..67H}. Cases 3a - 3d refer to a Saturn-sized planet and a super-Earth moon, cases 4a - 4d to a Neptune-sized planet and an Earth-sized moon, cases 5a - 5d to an Earth-Moon analog (though at either a 30\,d or a 58.31\,d orbital period around their star). In cases 6 - 9 we essentially redo all these cases except that the star resembles $\epsilon$\,Eri.

In cases 1a - 1p, we study the effect of the moon's orbital phase on the resulting transit shape and generate transits of planet-moon systems for orbital moon phases ranging from 0 to 1. We allow for different orbital inclinations of the moon orbit ($i_{\rm s}$) in these test cases (see Fig.~\ref{fig:diagram}), which may prevent occultations if the planet-moon orbit is sufficiently wide for a given inclination. As we show below, the inclination has an effect on the measured transit parameters if the line connecting the planet and the moon is parallel to the line of sight, that is, if the planet and the moon have a conjunction during the transit. For test cases 2a - 9d we choose the lowest inclination possible without causing a planet-moon occultation.

\subsection{Application to actual Kepler data}

As a first application of our search for exomoon indicators, we explore the time series of the fitted transit depth from \citet{2016ApJS..225....9H}. We then do a by-eye vetting of the transit depth time series, of the autocorrelation functions (ACF), and of the periodograms for each of the 2598 Kepler Objects of Interest (KOIs) listed in \citet{2016ApJS..225....9H} and identify KOI-1457.01 (Kepler-856\,b) as an interesting candidate. In brief, this is a roughly Saturn-sized validated planet with a false positive probability of $2.9~\times~10^{-5}$ \citep{2016ApJ...822...86M} and an orbital period of about 8\,d around an $\epsilon$\,Eri-like host star.

\section{Results}

\subsection{Signatures of exomoons in exoplanet-only transit model fits}

\begin{figure*}
    \centering
    \begin{overpic}[width=0.49\textwidth]{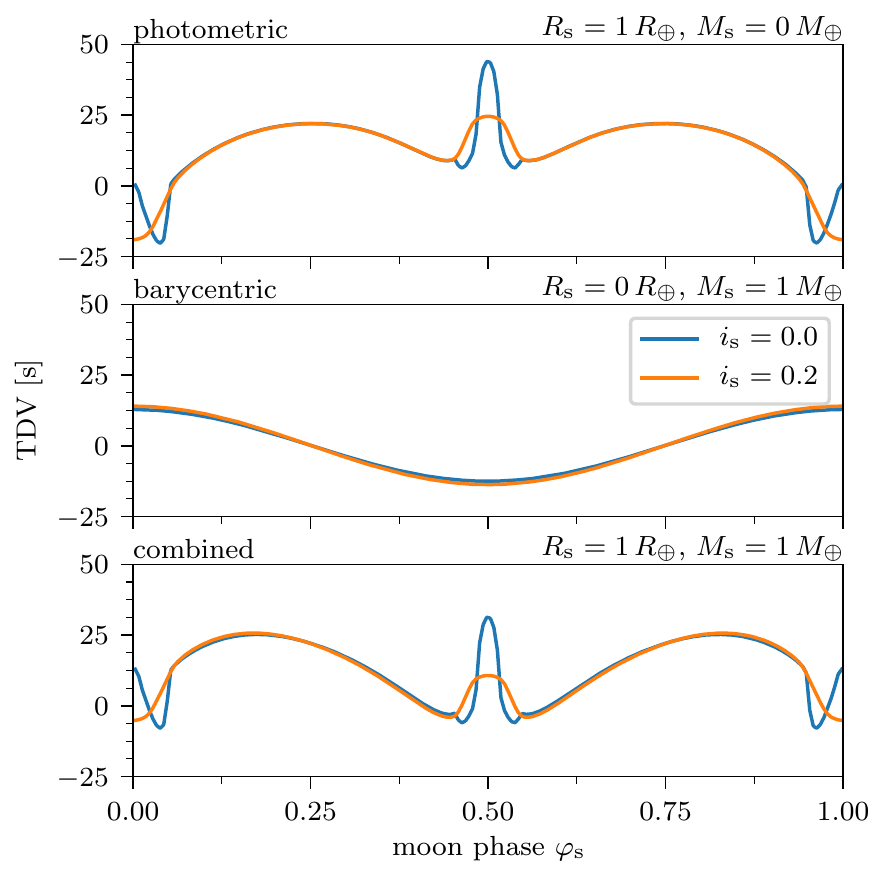}
 \put (2,87) {\Large$\displaystyle a)$}
\end{overpic}
   \begin{overpic}[width=0.49\textwidth]{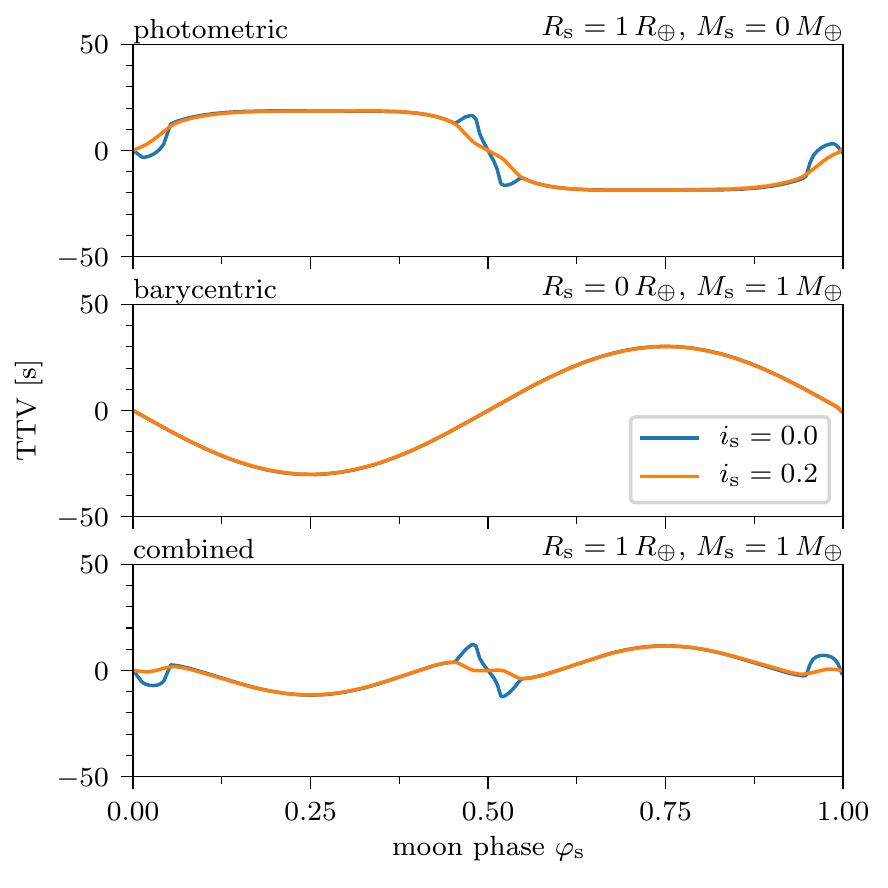}
 \put (2,87) {\Large$\displaystyle b)$}
\end{overpic}
\begin{overpic}[width=0.49\textwidth]{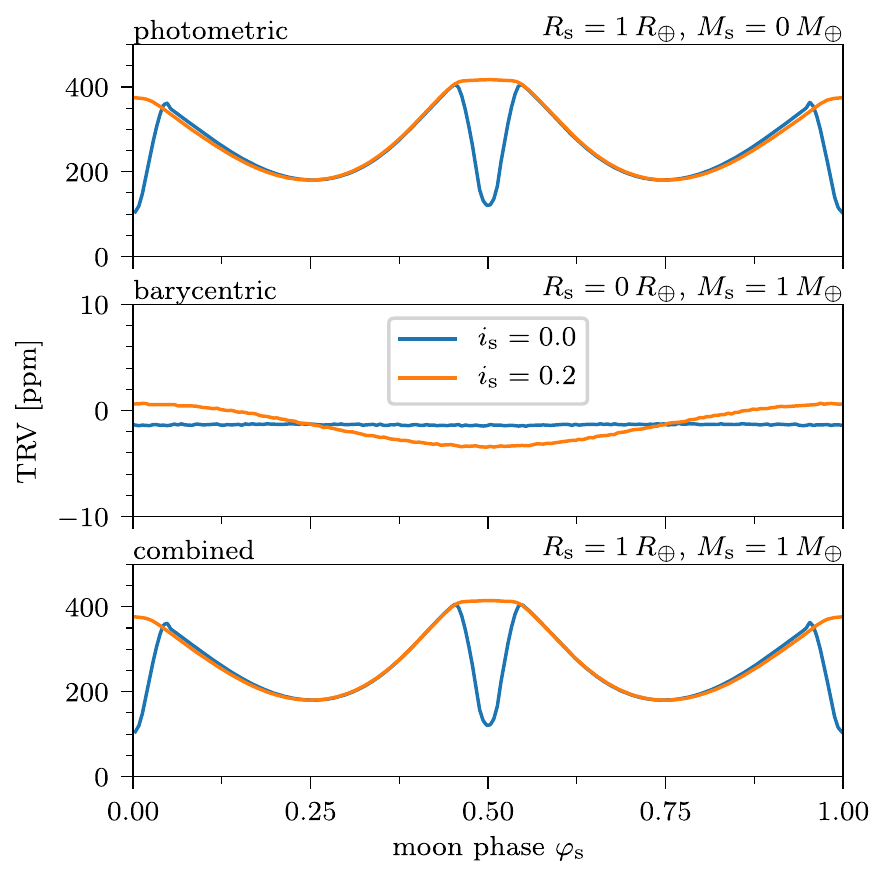}
 \put (2,87) {\Large$\displaystyle c)$}
\end{overpic}
\begin{overpic}[width=0.49\textwidth]{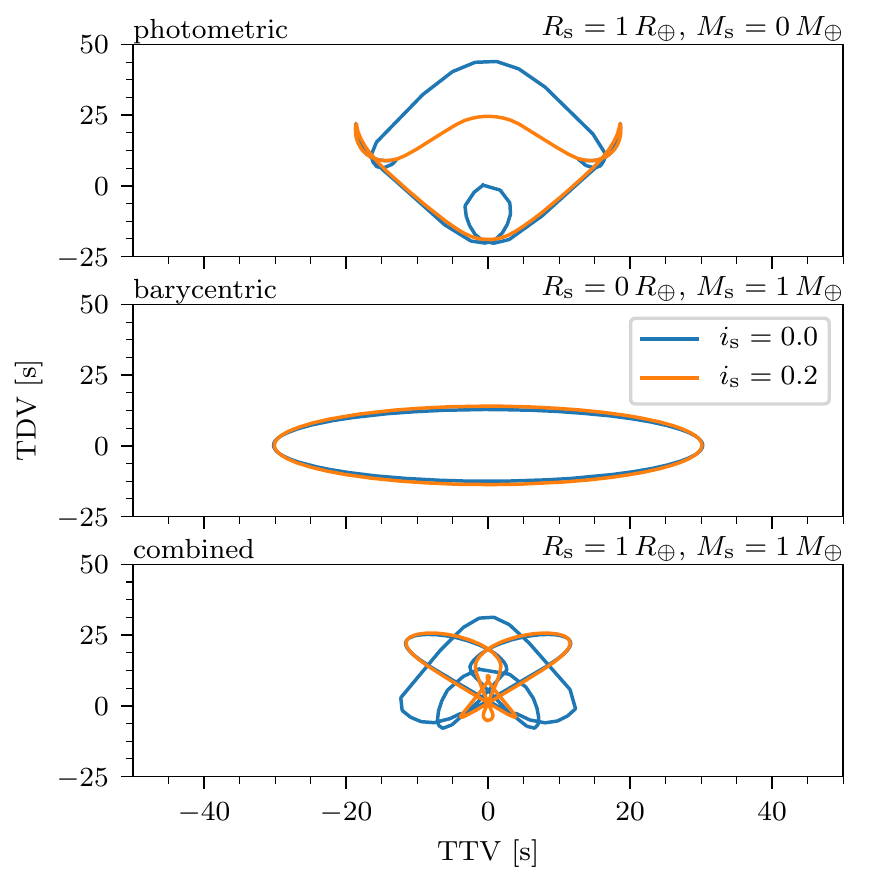}
 \put (2,87) {\Large$\displaystyle d)$}
\end{overpic}
    \caption{Exomoon indicators for a hypothetical Jupiter-Earth planet-moon system in a 30\,d orbit around a sun-like star as measured from noiseless simulated light curves by fitting a planet-only model. \textbf{(a)} Transit duration variation (TDV). \textbf{(b)} Transit timing variation (TTV). \textbf{(c)} Transit radius variation (TRV). \textbf{(d)} TTV vs. TDV diagram. Blue lines refer to a coplanar planet-moon system with $i_{\rm s}=0$ and orange lines depict the effects for an inclined planet-moon system with $i_{\rm s}=0.2$\,rad. In each panel, the top subplot shows the photometric contribution, the center subplot the barycentric contribution, and the bottom subplot the combined and measured effect.}
    \label{fig:TTV_TDV_TRV_photometric_barycentric_moon_phase}
\end{figure*}

Figure~\ref{fig:TTV_TDV_TRV_photometric_barycentric_moon_phase} shows the TDV (panel a), TTV (panel b), TRV (panel c), and combined TTV-TDV (panel d) effects. Each of these four panels is divided into three subplots showing the contributions of the photometric distortion of the light curve due to the moon (top subplots, moon mass set to zero), of the barycentric motion of the planet due to its moon (center subplots, moon radius set to zero), and the combined effect as derived from the planet-only fit to the simulated planet-moon transit light curve (bottom subplots). All panels assume a Jupiter-sized planet at an orbital period of $P_{\rm b}=30$\,d, an Earth-like moon with an orbital period of $P_{\rm s}=3.55$\,d ($=P_{\rm Eu}$), and a sun-like host star. Blue lines refer to an orbital inclination of $i_{\rm s}=0$, that is to say, to co-planar orbital configurations. This results in occultations near moon phases of 0, when the moon is in front of the planet as seen from Earth (see Fig.~\ref{fig:diagram}), and 0.5, when the planet is behind the moon. \mbox{Orange} lines refer to $i_{\rm s}=0.2$\,rad and do not produce planet-moon occultations.

The central panels of Fig.~\ref{fig:TTV_TDV_TRV_photometric_barycentric_moon_phase}, which illustrate the barycentric effects, correspond to the frameworks of \citet{1999A&AS..134..553S} (for the TTVs), \citet{2009MNRAS.392..181K} (for the TDVs), and \citet{2016A&A...591A..67H} (for the TTV-TDV figure). The contribution of the photometric distortion of the light curves by the moon (upper subplots), however, results in a very different fit of the TDV, TTV, TRV, and TTV-TDV figures (bottom subplots) than proposed in these papers. In particular, the fitted TTV-TDV figure does not resemble at all the ellipse that one would expect based on the barycentric contribution only. As a consequence, the TTV-TDV figure might not be as straightforward an exomoon indicator as proposed by \citet{2016A&A...591A..67H}.

\begin{figure*}
    \centering
    \includegraphics[width=0.49\textwidth]{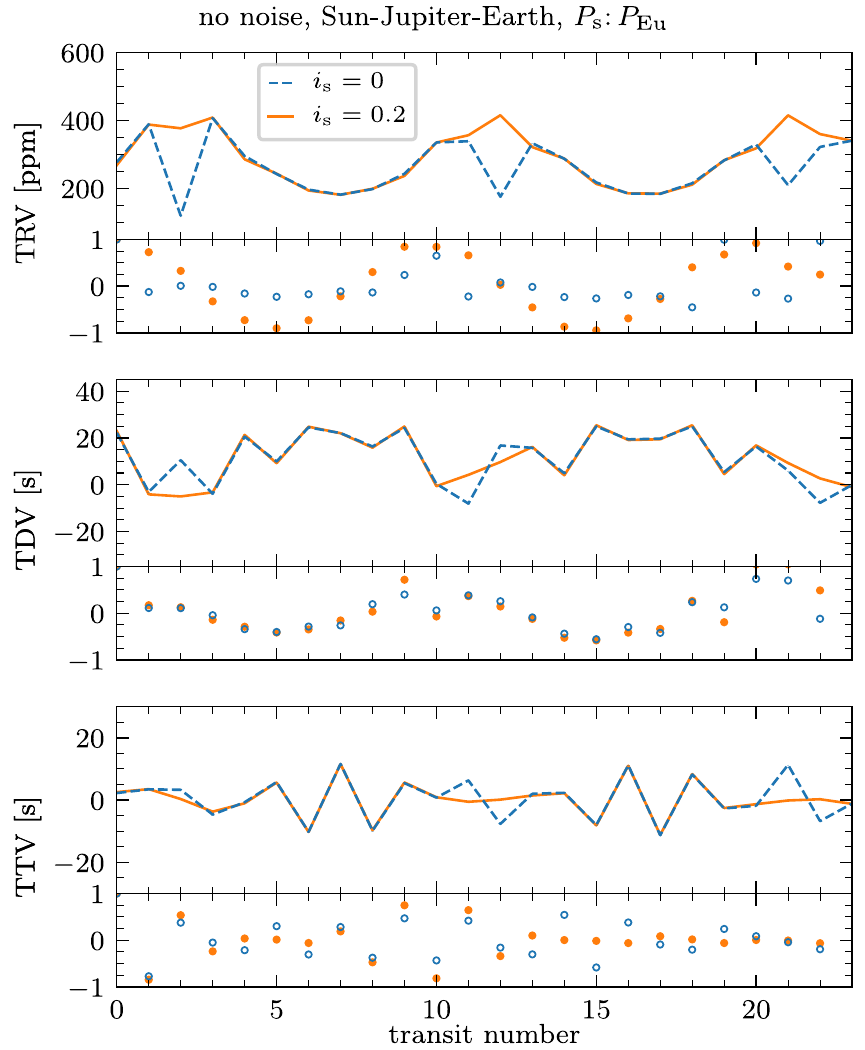}
    \includegraphics[width=0.49\textwidth]{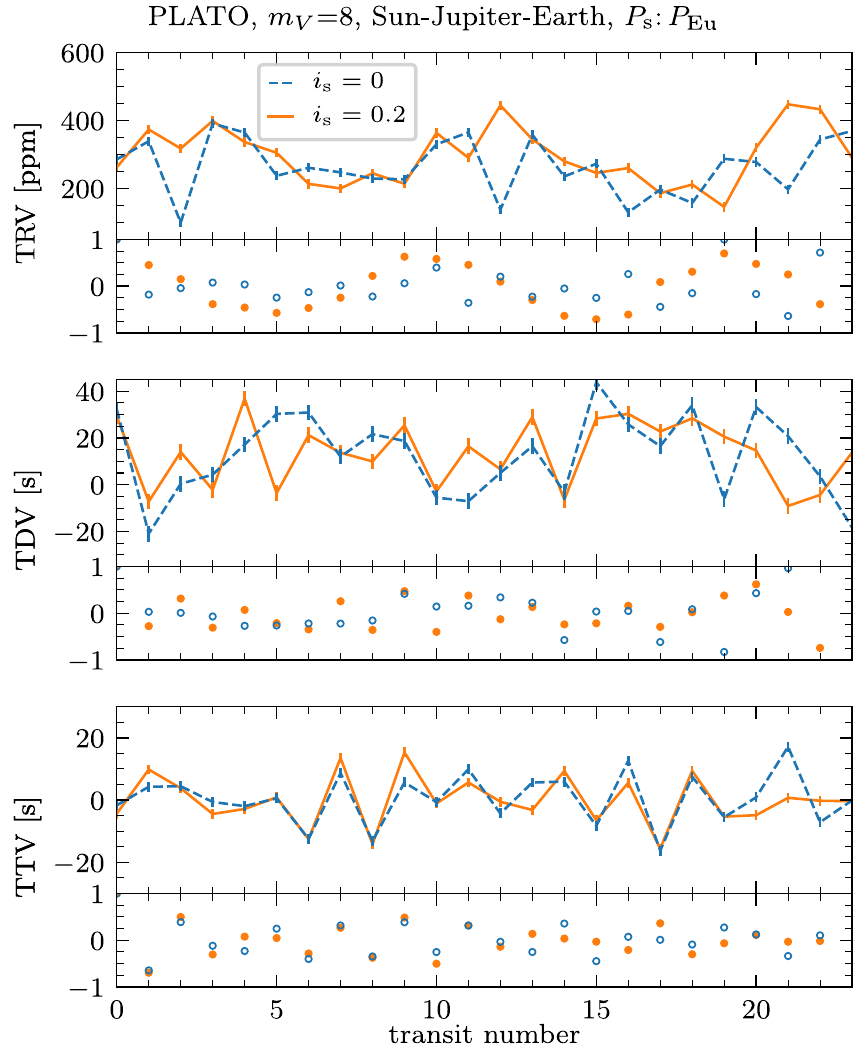}
    \caption{The TRV, TDV and TTV effects of an Earth-sized moon in a Europa-like orbit (3.55\,d) around a Jupiter-sized planet. The planet-moon barycenter has a 30\,d orbit around a sun-like star and the transits (not shown) were sampled with a PLATO-like cadence of 25\,s. \textit{Left:} Without noise. \textit{Right:} With noise contributions from stellar granulation and PLATO-like instrumental noise for an $m_V=8$ star. The subpanels show the autocorrelation. In all panels, blue symbols refer to a coplanar planet-moon system with $i_{\rm s}=0$ and orange symbols depict the effects for an inclined planet-moon system with $i_{\rm s}=0.2$\,rad.}
    \label{fig:Europa_comp_no_noise_and_PLATO}
\end{figure*}

The variation of the photometric TDV signal as a function of the moon's orbital phase ($0\leq\varphi_{\rm s}\leq1$) shows a relatively complicated behavior with minima when the planet and the moon align along the line of sight (moon phases of 0  or 1) and maxima at moon phases of 0.25, 0.5, and 0.75. For comparison, the variation of the barycentric TDV signal with changing moon phase is strongest when the planet and moon align at either a moon phase of 0 or 0.5. Both contributions combined, this leads to a complicated pattern for the TDVs as derived from the light curve fits.

The pattern of the photometric contribution to the measured TTV (Fig.~\ref{fig:TTV_TDV_TRV_photometric_barycentric_moon_phase}b) is shifted by half an orbital moon phase relative to the barycentric effect (panel a), though with roughly the same amplitude. The resulting TTV signal in the fitted model light curves then turns out to be extremely small (panel c).

The photometric TRV depends strongly on the projected separation of planet and moon, with $\varphi_{\rm s}~=~0$ corresponding to the moon being precisely in front of the planet as seen from Earth, $\varphi_{\rm s}~=~0.25$ corresponding to the moon being separated as much as possible from the planet and as ``late'' as possible for the transit etc. (see Fig.~\ref{fig:diagram}). The larger the sky-projected apparent separation between the planet and the moon, that is, the better the moon's photometric dip in the light curve is separated from the planet's photometric signature, the smaller the TRV. On the other hand, if the two bodies are sufficiently close to cause a planet-moon occultation, then the TRV drops significantly (see the dip at moon phase 0.5 in Fig.~\ref{fig:TTV_TDV_TRV_photometric_barycentric_moon_phase}c). In the case of an occultation there are two extreme scenarios to consider; one in which the occultation persists throughout the entire transit and one in which the occultation starts or ends during the transit. In the first case, the measured transit depth corresponds very closely to the true planet radius. In the second case, the resulting shape of the transit is strongly dependent on the exact orbital geometry and dynamics, and the inferred transit depth can differ strongly from the planet-only scenario (see Fig.~\ref{fig:TTV_TDV_TRV_photometric_barycentric_moon_phase}c).

Interestingly, there is a difference in amplitude depending on whether the moon passes behind or in front the planet. This is due to the slightly different transit shapes of the combined planet-moon system. If the moon's effective (tangential) speed across the stellar disk is lower than that of the planet, the overall transit shape resembles that of a single planet except for an additional signature in the wings of the transit light curve. Note that this asymmetry of the TRV effect cannot help to discriminate between prograde and retrograde moon orbits, which is both a theoretical and an observational challenge for exomoon characterization \citep{2014ApJ...791L..26L,2014ApJ...796L...1H}. Moreover, in practice this effect will hardly be observable when noise is present. Note that the barycentric TRV shows no variation as a function of the moon phase in the case of $i_{\rm s}=0$ but some variation in the case of $i_{\rm s}=0.2$\,rad. This can be explained by the planet crossing the star at a higher or lower impact parameter due to its movement around the planet-moon barycenter. At different transit impact parameters then, the resulting transit depth will be different due to the stellar limb darkening effect \citep{2019A&A...623A.137H}.

\subsection{TTV, TDV, and TRV from PLATO-like light curves}

The panels at the left of Fig.~\ref{fig:Europa_comp_no_noise_and_PLATO} show the TRV (top), TDV (center), and TTV (bottom) effects, respectively, as measured from a sequence of simulated light curves without noise. Just like in Fig.~\ref{fig:shape-shift}, we modeled the planet moon transits as such but fitted a planet-only model. The number of the transit in this sequence is shown along the abscissa. Also shown are the respective autocorrelations below each of these three time series. The model systems used for these sequences are our test cases 1b ($i_{\rm s}=0$) and 1j ($i_{\rm s}=0.2$\,rad), that is a Jupiter-sized, Jupiter-mass planet with an Earth-sized moon in a 3.55\,d (Europa-like) orbit. The center of mass of this planet-moon binary is in a 30\,d orbit around a sun-like star (for details, see Table~\ref{tab:params}).

The combination of these particular orbital periods around the star ($P_{\rm b}$) and around the planet-moon barycenter ($P_{\rm s}$) happen to make the moon jump by a phase of 0.448 circumplanetary (or circumbarycentric) orbits between successive transits. This well-known undersampling results in an observed TTV signal period of under 2 $P_{\rm b}$ \citep{2009MNRAS.392..181K}. The TDV series is also subject to this aliasing effect, but the autocorrelation suggests some periodicity. Finally, the TRV signal, whose period is just a half of $P_{\rm s}$, is not affected by the undersampling. Hence, the periodicity can be observed even by eye in the top panel and, of course, also in the corresponding autocorrelation.

Moving on to the addition of noise in our simulated light curves, the panels at the right of Fig.~\ref{fig:Europa_comp_no_noise_and_PLATO} illustrate the same TRV (top), TDV (center), and TTV (bottom) series as in the panels at the left, but now as if the host star were an $m_V=8$ solar type star observed with PLATO. In both the TDV and the TTV panels we see that the magnitude of the error caused by the noise in the light curves is comparable to the TTV and TDV amplitudes. As a consequence, both the two time series and their autocorrelations show only weak hints of periodic variations. For comparison, the TRV signal of the slightly inclined system ($i=0.2$\,rad, orange dashed line) is still clearly visible, whereas the signal for the $i=0$ system is somewhat weaker.

\begin{figure*}
    \centering
    \includegraphics[width=.49\textwidth]{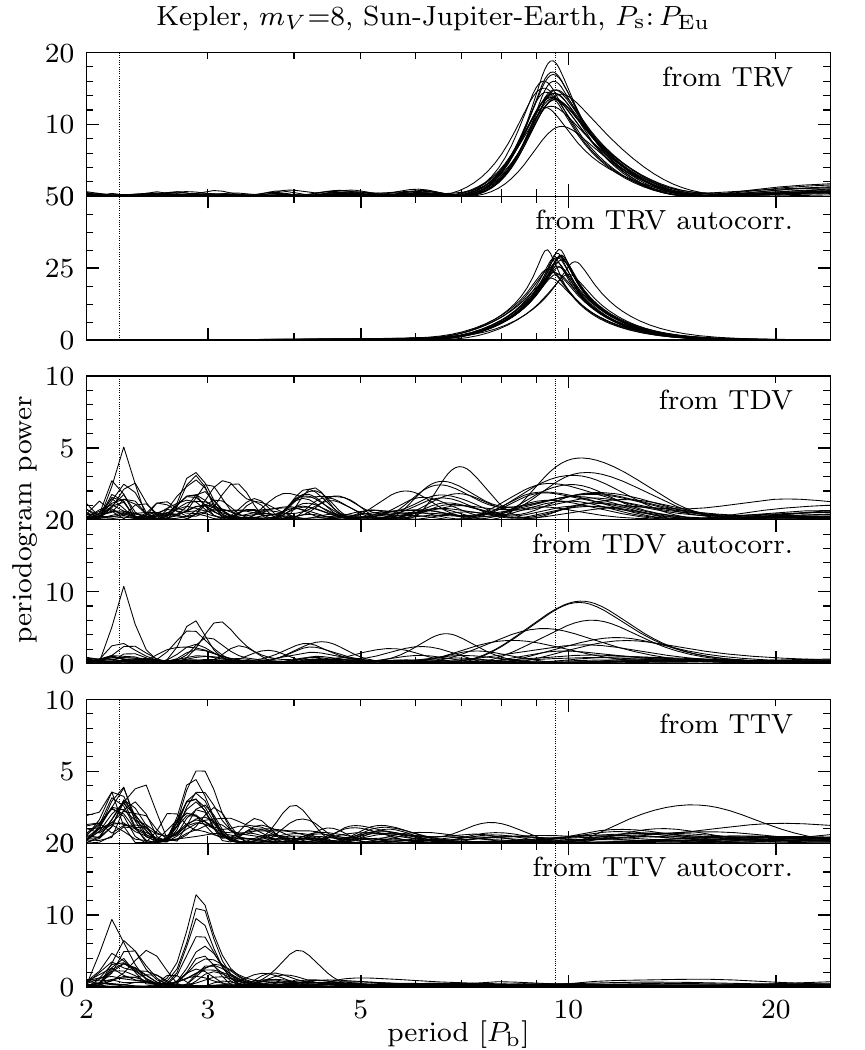}
    \includegraphics[width=.49\textwidth]{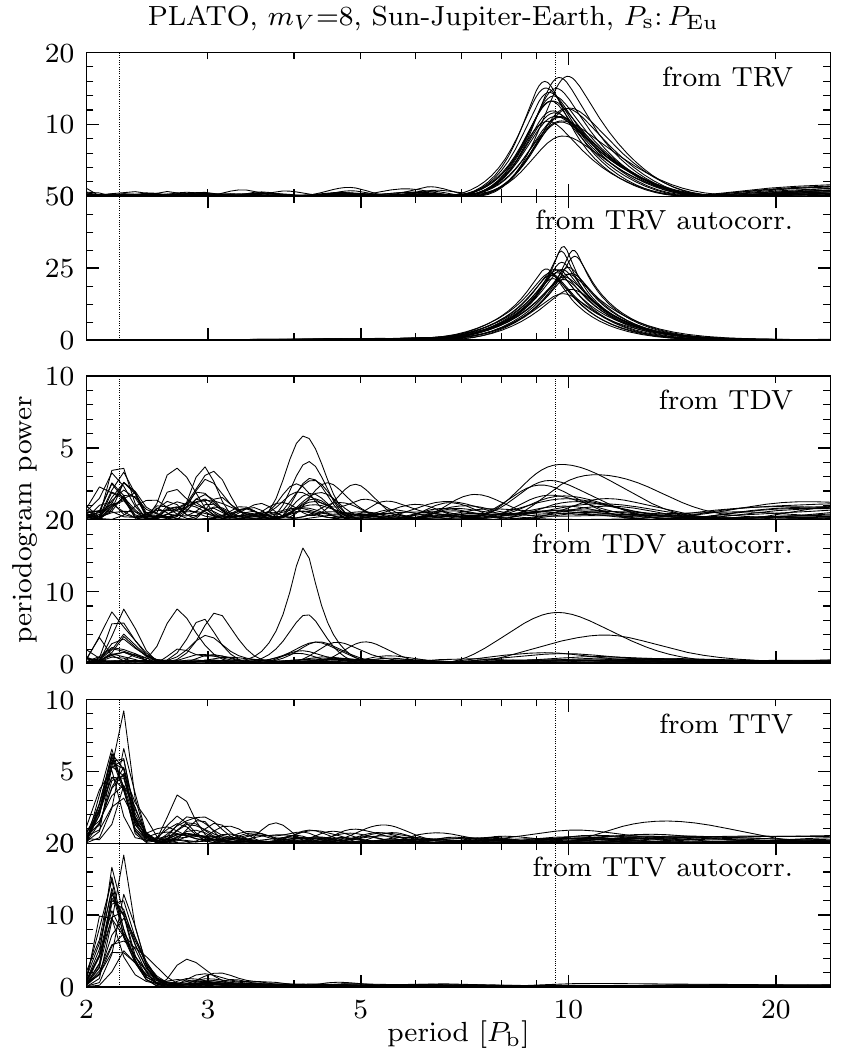}
    \caption{Comparison of the simulated performance of Kepler (left) and PLATO (right) in detecting the TRV, TDV, and TTV effects. A sequence of 20 simulated light curves for a Jupiter-Earth planet-moon system around a sun-like star were pre-computed and analyzed, similar to the system used in Fig.~\ref{fig:Europa_comp_no_noise_and_PLATO} (test case 1p in Table~\ref{tab:params}). The 20 randomized noise realizations contain both white noise and granulation equivalent to an $m_V=8$ star. The Nyquist period of the system is equal to $2\,P_{\rm b}$. The panels shows the periodograms of the TRV (top), TDV (center), and TTV (bottom) effects and their respective autocorrelations.}
    \label{fig:shifted_noVIRGO_jupiter_earth}
\end{figure*}

To obtain a more quantitative understanding of the possible periodicities in the TRV, TTV, and TDV sequences and in order to compare the performances of the Kepler and PLATO missions, we investigate the periodograms of the data. In Fig.~\ref{fig:shifted_noVIRGO_jupiter_earth} we present the periodograms of the time series and of the autocorrelations of the TRV (top), TTV (center), and TDV (bottom) sequences. The left column of panels assumes a S/N performance and 30\,min cadence akin to the Kepler mission, whereas the right column refers to a 25\,s sampling and coverage by all 24 of PLATO's normal cameras. Each panel is based on 20 randomized noise realizations of 24 transits of the same system as in Fig.~\ref{fig:Europa_comp_no_noise_and_PLATO}. In both the Kepler and the PLATO case, the TRVs of each realization show a clear peak at around 9.6 times the orbital period of the barycenter around the star. This peak occurs in the periodograms of both the time series and of the autocorrelation of the time series and can be explained as an aliasing effect (see Appendix~\ref{sec:aliasing} for a calculation). The periodograms of the TDV series and TDV autocorrelation, however, do not show any significant signals near this period, although the TTVs show a peak in the periodogram close to the period corresponding to the Nyquist frequency.

Of similar interest is our finding that Kepler and PLATO have comparable performances in detection TRVs and TDVs, but PLATO data produces a more significant signal in the TTV periodogram and autocorrelation. This result is mostly due to the higher cadence of PLATO. What is more, PLATO will explicitly observe bright sun-like stars, which means that this type of data (whether with or without moons) will actually be available. For comparison, the brightest exoplanet host star observed with Kepler, HD\,179070 (Kepler-21), has an $m_V$ magnitude of 8.3 \citep{2012ApJ...746..123H} and most Kepler host stars are substantially fainter. As a consequence, any transiting planet with an exomoon in the Kepler data would yield a signal that is significantly weaker than the one shown in Fig.~\ref{fig:shifted_noVIRGO_jupiter_earth}.

\subsection{The TRV effect of different star-planet-moon systems}
\label{sec:cases2to9}

\begin{figure*}
    \centering
    \includegraphics[width=.99\textwidth]{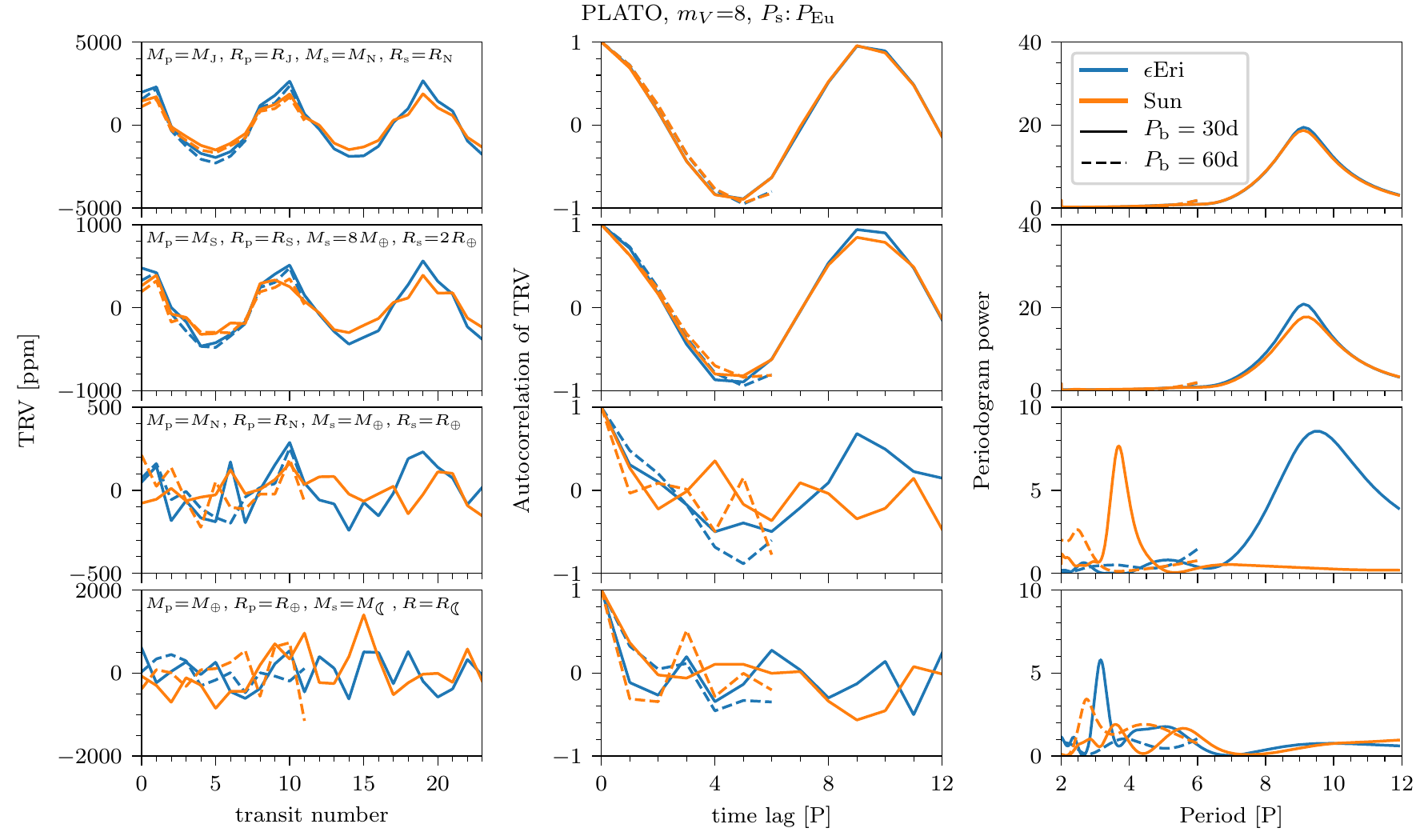}
    \caption{Transit sequence of the transit radius variation (TRV, left), autocorrelation function of the TRV transit sequence (center), and periodogram of the autocorrelation (right) for various planet-moon configurations around an $m_{V}=8$ star as observed with PLATO (see Sect.~\ref{sec:cases2to9}). Blue lines refer to an $\epsilon$\,Eridani-like star, orange lines to a sun-like star. Subscripts ``J'', ``S'', ``N'', ``$\oplus$'', and ``$\leftmoon$'' for the radii and masses refer to Jupiter, Saturn, Neptune, Earth, and the Moon, respectively. In the Jupiter-Neptune and Saturn/Super-Earth cases the TRV signal is clearly visible in the measured TRV time series. Consequently the signal is also visible in the autocorrelation. Due to only calculating the AC up to a time lag of have the observation length, a peak in the periodograms is only visible in the $P_{\rm b}=30~$d cases and not the $P_{\rm b}=60~$d cases. In the Neptune-Earth cases the signal in the TRV time series is barely visible, while the autocorrelation and periodogram for the $P_{\rm b}=30~$d cases show a clear signal. There is no visible signal in any of the Earth-Moon cases for in the corresponding time series, autocorrelation, and periodogram.}
    \label{fig:combined_systems_all_metrics}
\end{figure*}

The results presented in Figs.~\ref{fig:Europa_comp_no_noise_and_PLATO} and \ref{fig:shifted_noVIRGO_jupiter_earth} refer to only two specific test cases, that is, 1b and 1f, respectively. Next, we extend our investigations to all other test cases described in Sect.~\ref{sec:parameterization} and summarized in Table~\ref{tab:params}. Appendix~\ref{sec:plots_plots_plots} contains a completed set of plots for each of the subcases, while Fig.~\ref{fig:combined_systems_all_metrics} shows the results for a subsection of these cases as a cut through the parameter space. All results shown in Fig.~\ref{fig:combined_systems_all_metrics} assume synchronous observations of an $m_V=8$ star with all 24 normal cameras of PLATO. We focus on the TRV because our investigations to this point have indicated that the TRV effect is more pronounced than the TTV and TDV effects.

The four systems in the upper row in Fig.~\ref{fig:combined_systems_all_metrics} refer to cases 2b (solid orange), 2d (dashed orange), 6b (blue solid), and 6d (blue dashed). The second line of panels shows cases 3b (solid orange), 3d (dashed orange), 7b (solid blue) and 7d (dashed blue). The third line of panels shows cases 4b (solid orange), 4d (dashed orange), 8b (blue solid), and 8d (blue dashed). The bottom row of panels refers to cases 5b (orange solid), 5d (orange dashed), 9b (blue solid), and 9d (blue dashed).

In cases 2, 3, 6, and 7 involving the large Neptune and super-Earth-sized moons, the TRV signal is always visible in the time series (left column). The autocorrelation function of the TRV time series (center column) for systems with $P_{\rm b}=60$\,d, however, is truncated at 6 times the orbital period due to the low number of transits available at that period. This truncation is caused by our our restriction of the autocorrelation function to a lag time of at least half the observed number of transits. For comparison, in those cases that assume $P_{\rm b}=30$\,d and, thus, twice the number of transits available for the analysis, the signal in the corresponding periodogram of the autocorrelation is very clear.

For cases 4 and 8 involving an Earth-sized moon, the detectability of the TRV signal depends on the moon's orbital period and the size of its star. For the case of a Europa-like orbital period of the moon and the $\epsilon$\,Eridani-like star, a marginally significant peak in the periodogram is present, but no significant peak can be seen if the star is sun-like or the moon's orbital period is that of Io. In none of the cases with a Moon-sized moon (cases 5 and 9) we were able to produce a significant TRV signal.

In the right column of panels in Fig.~\ref{fig:combined_systems_all_metrics}, we observe that our test moons with a Europa-like orbital period all cause a peak in the periodograms at around $9.6\,P_{\rm b}$. For all the other cases involving a moon with an Io-like period (all subcases a and c), the position of the peak in the periodgrams of the autocorrelation is near $12\,P_{\rm b}$, which is at the edge of the number of orbital periods we deem reasonable for analysis.

The positions of the peak in the periodogram of the TRV autocorrelation function corresponds to the remainder of the fraction between the circumstellar period of the planet and the circumplanetary period of the moon. If the nominal value of this resulting peak position is smaller than two planetary orbital periods, then the actual position of the peak is shifted to a position longer than two times $P_{\rm b}$ (see Appendix~\ref{sec:aliasing}).

\subsection{Kepler-856\,b as an example for TRVs}

\begin{figure*}
    \centering
    \includegraphics[width=1.01\textwidth]{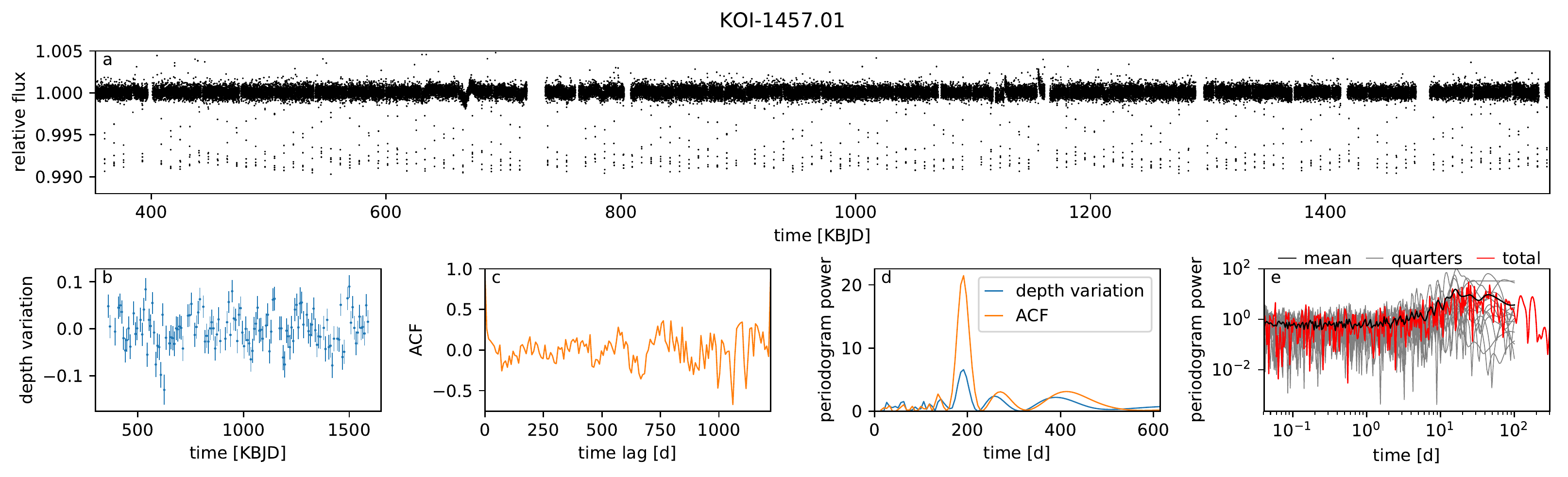}
    \caption{Analysis of the Kepler light curve of Kepler-856\,b (KOI-1457.01). \textbf{(a)} PDCSAP Kepler light curve. \textbf{(b)} Fractional transit depths variation from \citet{2016ApJS..225....9H} for a total of 134 transits. \textbf{(c)} Autocorrelation of the transit depth variation. \textbf{(d)} Periodogram of the transit depth variation and periodogram of the autocorrelation function. \textbf{(e)} Periodograms of the PDCSAP flux of 14 Kepler quarters (gray lines) and their mean (black line) after masking out the planetary transits.}
    \label{fig:TRVs_Kepler856b}
\end{figure*}

As a first application of our novel TRV method, we searched for TRVs in the published values of the transit depth sequences of 2598 Kepler Objects of Interest (KOIs), or transiting exoplanet candidates, from \cite{2016ApJS..225....9H}. As a general result of our by-eye inspection of those KOIs with the highest periodogram power, we noticed that stellar rotation is the major cause of false positives. In the case of Kepler-856\,b (KOI-1457.01), however, the interpretation of a false positive is less evident based purely on the inspection of the light curve. Figure~\ref{fig:TRVs_Kepler856b} shows our vetting sheet constructed for our TRV survey with Kepler. Kepler-856\,b is a Saturn-sized planet with unknown mass in a 8\,d orbit around a sun-like star. The transit sequence has been used to statistically validate Kepler-856\,b as a planet with less than 1\,\% probability of being a false positive \citep{2016ApJ...822...86M}.

Figure~\ref{fig:TRVs_Kepler856b}(a) shows the Pre-Search Conditioning Simple Aperture Photometry (PDCSAP) Kepler flux \citep{2010ApJ...713L..87J}, which has been corrected for the systematic effects of the telescope, normalized by the mean of each Kepler quarter. We note that the apparent periodic variation of the maximum transit depth is probably not the TRV effect that we describe. The apparent variation of the maximum transit depth visible in Fig.~\ref{fig:TRVs_Kepler856b}(a) is likely an aliasing effect due to the finite exposure time of the telescope.

Figure~\ref{fig:TRVs_Kepler856b}(b) shows the series of the transit depths of Kepler-856\,b, panel (c) the autocorrelation function, panel (d) the periodogram power of the transit depth series (blue) and of the autocorrelation function (orange), and panel (e) shows the periodograms of the 14 Kepler quarters of the light curve (gray) together with their mean values (black). We used panel (e) as a means to quickly identify stellar variability and compare it to any possible periodicity in the TRV periodogram of panel (d).

An interesting observation of Fig.~\ref{fig:TRVs_Kepler856b}(d) that caught our attention is the striking peak in the periodogram of the autocorrelation function near 200\,d and, to a minor extent, a peak in the periodogram of the transit depth variation at the same period. These observed features are reminiscent of the features in the corresponding periodograms of the simulated data shown in Figs.~\ref{fig:combined_systems_all_metrics}, \ref{fig:combined_systems_all_metrics_magn_8}, and \ref{fig:combined_systems_all_metrics_magn_11}, which are caused by large moons. Moreover, the periodogram of the total light curve, which is a combination of the all Kepler quarters (red line in Fig.~\ref{fig:TRVs_Kepler856b}e), does not show a peak near 200\,d with similar amplitude as the periodograms in panel (d), although there are two moderate peaks between 100\,d and 200\,d. We conclude that there is only marginal evidence for stellar activity as the origin of the TRV signal in panel (d).

But could a moon actually be stable around Kepler-856\,b from an orbital dynamics point of view? Using the stellar and planetary parameters derived by \citet{2016ApJ...822...86M}, we estimate that the planetary Hill radius ($R_{\rm H}$) is about 540,000\,km wide. Given that prograde moons can only be stable out to about $0.5\,{\times}\,R_{\rm H}$ \citep{2006MNRAS.373.1227D}, the range of prograde and gravitationally stable moon orbits is limited to about 270,000\,km, corresponding to about $9\,R_{\rm p}$ for Kepler-856\,b. For comparison, Io, the innermost of the Galilean moons, orbits Jupiter at about 6.1 planetary radii, corresponding to about 8\,\% of Jupiter's Hill radius. So the space for orbital stability is quite narrow around Kepler-856\,b compared to the orbits of the solar system moons, though a close-in sufficiently large moon could still be physically plausible. 

\section{Discussion}

Comparing the TTV, TDV, and TRV distributions as a function of the moon's orbital phase for systems of various orbital inclinations, we noticed that systems without occultations (and significant inclinations) exhibit more sinusoidal patterns. A comparison of the blue (with occultations) and orange (without occultations) lines in any of the upper panels of Fig.~\ref{fig:TTV_TDV_TRV_photometric_barycentric_moon_phase}(a) - (c) serves as an illustrative example for this. Configurations without occultations thus naturally produce higher peaks in in the corresponding periodograms. An improvement to our TRV indicator, in particular for systems with occultations, could be achieved if the TRV signal could be modeled and fitted to the TRV time series, or if it could be used as a fitting function in the periodogram. This approach, however, could become as mathematically and computationally involved as implementing a planet-moon transit model to the light curve to begin with.

The detectability of moon-induced TRV signals increases with the number of transits observed and it is desirable to at least observe a full period of the TRV signal. Hence, this new exomoon indicator can preferably be detected with longterm transit surveys such as Kepler or PLATO, which deliver (or are planned to deliver) continuous stellar high-precision photometry for several years.

Localized features on a star's surface such as spots or faculae can also cause variations in the measured exoplanet radius. Hence, any search for moon-induced TRVs needs to include the analysis of stellar variations and verify that any candidate signals are not caused by the star. Analyzing the contributions of the barycentric and of the photometric components to the observed -- or in our case the fitted -- TTV signal, we noticed that both components tend to cancel each other, see Fig.~\ref{fig:TTV_TDV_TRV_photometric_barycentric_moon_phase}(b). This means that for low-density moons the TTV signal will be dominated by the photometric component, while for high-density moons the barycentric component will be more relevant. For moons with densities similar to the densities of the terrestrial or planets and moons of the solar system, which range between about $3\,{\rm g\,cm}^{-3}$ and $5\,{\rm g\,cm}^{-3}$, the overall TTV effect might actually be very hard to observe as the barycentric and photometric and contributions essentially cancel each other out. 

The amplitude of the TRV signal increases with increasing orbital separation between the planet and the moon. The TRV amplitude is maximized when the orbital semimajor axis around the planet is larger than two stellar radii, at which point the two transits of the planet and the moon can occur without any overlap in the light curve. Transits without an overlap happen when the moon's orbital phase is near $\varphi_{\rm s}=0.25$ or $\varphi_{\rm s}=0.75$, that is, when the planet and the moon show maximum tangential deflection on the celestial plane. In these cases of a transit geometry, there is no ``contamination'' of the planetary transit by the moon and the lowest possible transit radius is fitted, thereby maximizing the variation with respect to those transits of a transit sequence in which the planet and the moon transit along or near the line of sight, that is, near $\varphi_{\rm s}=0$ or $\varphi_{\rm s}=0.5$, at which point the maximum transit radius (with maximum contamination by the moon) is fitted.

\section{Conclusions}

We use standard tools for the analysis of exoplanet transits to study the usefulness of TTVs, TDVs, and TRVs as exomoon indicators. Our aim is to identify patterns in the data that is now regularly being obtained by exoplanet searches, which could betray the presence of a moon around a given exoplanet without the need of dedicated star-planet-moon simulations.

We first test these indicators on simulated light curves of different stellar systems and find that the TRVs observed over multiple transits show periodic patterns that indicate the presence of a companion around the planet. The TTV and TDV signals, however, may be less useful indicators for the presence of an exomoon, depending on whether the density of the moon. The moon density determines whether the barycentric or photometric component of the respective effect dominates. The TRV signal, for comparison, is always dominated by the photometric contribution and, as a consequence, has more predictable, understandable variation.

Our search for indications of an exomoon in the 2598 series of transit mid-points, transit durations, and transit depths by \citet{2016ApJS..225....9H} reveals the hot Saturn Kepler-856\,b as an interesting object with strong TRVs that are likely not caused by stellar variability. While an exomoon interpretation would be a natural thing to put forward in this study, we caution that the range of dynamically stable moon orbits around Kepler-856\,b is rather narrow due to its proximity to the star. The planetary Hill sphere is about $18\,R_{\rm p}$ wide, restricting prograde moons to obits less than about $9\,R_{\rm p}$ wide. At this proximity to the planet, however, tides may become important and ultimately present additional obstacles to the longterm survival of a massive moon \citep{2002ApJ...575.1087B,2012A&A...545L...8H}. Generally speaking, our investigations of the fitted transit depths from \citet{2016ApJS..225....9H} and of the corresponding Kepler light curves show that rotation of spotted stars or an undersampling effect of the transit light curve can cause false positives.

The previously suggested method of finding exomoons in the TTV-TDV diagram \citep{2016A&A...591A..67H} might not be as efficient a tool to search for moons as thought. The TTV-TDV diagram only forms an ellipse in the barycentric regime, that is, when the moon has a very high density. For rocky or icy moons, however, the figure of the TTV-TDV diagram derived from fitting a planet-only model to the combined planet-moon transit light curve is much more complex than an ellipse.

We find that the threshold for a TRV-based exomoon detection around an $m_V=8$ solar type star with the Kepler or PLATO missions lies at a minimum radius of about the size of the Earth. Our simulations suggest that PLATO and Kepler have a similar performance in the detection of the exomoon-induced TRV effect. That said, Kepler did not actually observe any transiting planet around a star as bright as $m_V=8$, whereas PLATO will explicitly target these stars for transiting planets. Moreover, we find that PLATO has a better chance of detecting exomoon-induced TTVs due to its higher cadence.

\begin{acknowledgements}
This work was supported in part by the German Aerospace Center (DLR) under PLATO Data Center grant 50OL1701.
\end{acknowledgements}

\bibliographystyle{aa} 
\bibliography{ms}

\appendix

\section{Full parameter set}

\setlength{\rotFPtop}{0.\textwidth}
\begin{sidewaystable}
\tiny
\setlength{\tabcolsep}{1.5pt}
    \centering
    
    \begin{tabular}{c|cccccccccccccccc}
          &  \multicolumn{8}{c}{Case ID}\\
         & 1a & 1b & 1c & 1 d & 1e & 1f & 1g & 1h & 1i & 1j & 1k & 1l & 1m & 1n & 1o & 1p\\
        \hline\hline 
        Parameter & \multicolumn{16}{c}{ }\\
        \hline
        $R_\star$ [$R_\odot$]& $1$ & $1$ & $1$ & $1$ & $1$ & $1$ & $1$ & $1$ & $1$ & $1$ & $1$ & $1$ & $1$ & $1$ & $1$ & $1$\\
        $M_\star$ [$M_\odot$]& 
        $1$ &$1$ & $1$ & $1$ & $1$ &$1$ & $1$ & $1$ & $1$ &$1$ & $1$ & $1$ & $1$ &$1$ & $1$ & $1$\\
        $q_1$  & 0.6& 0.6 & 0.6 & 0.6 & 0.6 & 0.6 & 0.6 & 0.6 & 0.6& 0.6 & 0.6 & 0.6 & 0.6 & 0.6 & 0.6 & 0.6\\
        $q_2$  & 0.4 & 0.4 & 0.4 & 0.4 & 0.4 & 0.4 & 0.4 & 0.4 & 0.4 & 0.4 & 0.4 & 0.4 & 0.4 & 0.4 & 0.4 & 0.4\\
        \hline
        $R_{\rm p}$ & 
        $R_{\rm J}$ & $R_{\rm J}$ & $R_{\rm J}$ & $R_{\rm J}$ & $R_{\rm J}$ & $R_{\rm J}$ & $R_{\rm J}$ & $R_{\rm J}$& 
        $R_{\rm J}$ & $R_{\rm J}$ & $R_{\rm J}$ & $R_{\rm J}$ & $R_{\rm J}$ & $R_{\rm J}$ & $R_{\rm J}$ & $R_{\rm J}$\\
        $M_{\rm p}$  & 
        $M_{\rm J}$ & $M_{\rm J}$ & $M_{\rm J}$ & $M_{\rm J}$ & $M_{\rm J}$ & $M_{\rm J}$ & $M_{\rm J}$ & $M_{\rm J}$& 
        $M_{\rm J}$ & $M_{\rm J}$ & $M_{\rm J}$ & $M_{\rm J}$ & $M_{\rm J}$ & $M_{\rm J}$ & $M_{\rm J}$ & $M_{\rm J}$\\
        $P_{\rm b}$ [d]& $30$ & $30$ & 30 & 30& $30$ & $30$ & 30 & 30 & $30$ & $30$ & 30 & 30& $30$ & $30$ & 30 & 30\\
        $b_{\rm b}$  & 0.1 & 0.1 & 0.1 & 0.1 & 0.1 & 0.1 & 0.1 & 0.1 & 0.1 & 0.1 & 0.1 & 0.1 & 0.1 & 0.1 & 0.1 & 0.1\\
        $\varphi_{\rm b}$  & 0 & 0 & 0 & 0 & 0 & 0 & 0 & 0 & 0 & 0 & 0 & 0 & 0 & 0 & 0 & 0\\
        \hline
        $R_{\rm s}$  & 0 & 0 & 0 & 0 & $R_\oplus$ & $R_\oplus$ & $R_\oplus$ & $R_\oplus$ & 0 & 0 & 0 & 0 & $R_\oplus$ & $R_\oplus$ & $R_\oplus$ & $R_\oplus$\\
        $M_{\rm s}$  & 0 & 0 & $M_\oplus$ & $M_\oplus$ & 0 & 0 & $M_\oplus$ & $M_\oplus$ & 0 & 0 & $M_\oplus$ & $M_\oplus$ & 0 & 0 & $M_\oplus$ & $M_\oplus$\\
        $P_{\rm s}$  & $P_{\rm Io}$ & $P_{\rm Eu}$ & $P_{\rm Io}$ & $P_{\rm Eu}$ & $P_{\rm Io}$ & $P_{\rm Eu}$ & $P_{\rm Io}$ & $P_{\rm Eu}$ & $P_{\rm Io}$ & $P_{\rm Eu}$ & $P_{\rm Io}$ & $P_{\rm Eu}$ & $P_{\rm Io}$ & $P_{\rm Eu}$ & $P_{\rm Io}$ & $P_{\rm Eu}$\\
        $\varphi_{\rm s}$  & 0-1 & 0-1 & 0-1 & 0-1 & 0-1 & 0-1 & 0-1 & 0-1 & 0-1 & 0-1 & 0-1 & 0-1 & 0-1 & 0-1 & 0-1 & 0-1\\
        $i_{\rm s}$ & 0 & 0 & 0 & 0 & 0 & 0 & 0 & 0 & 0.2 & 0.2 & 0.2 & 0.2 & 0.2 & 0.2 & 0.2 & 0.2  
    \end{tabular}
    \begin{tabular}{c|cccccccccccccccc|ccccccccccccccccc}
          &  \multicolumn{32}{c}{Case ID}\\
         & 2a & 2b & 2c & 2d & 3a & 3b & 3c & 3d & 4a & 4b & 4c & 4d & 5a & 5b  & 5c & 5d & 6a & 6b & 6c & 6d & 7a & 7b & 7c & 7d & 8a & 8b & 8c & 8d & 9a & 9b & 9c & 9d\\
        \hline\hline 
        Parameter & \multicolumn{32}{c}{ }\\
        \hline
        $R_\star$ [$R_\odot$]& $1$ & $1$ & $1$ & $1$ & $1$ & $1$ & $1$ & $1$ & $1$ & $1$ & $1$ & $1$ & $1$ & $1$ & $1$ & $1$ & $0.85$ & $0.85$ & $0.85$ & $0.85$ & $0.85$ & $0.85$ & $0.85$ & $0.85$ & $0.85$ & $0.85$ & $0.85$ & $0.85$ & $0.85$ & $0.85$ & $0.85$ & $0.85$\\
        $M_\star$ [$M_\odot$]&
        $1$ &$1$ & $1$ & $1$ & $1$ &$1$ & $1$ & $1$ & $1$ &$1$ & $1$ & $1$ & $1$ &$1$ & $1$ & $1$ & $0.84$ & $0.84$ & $0.84$ & $0.84$ & $0.84$ & $0.84$ & $0.84$ & $0.84$ & $0.84$ & $0.84$ & $0.84$ & $0.84$ & $0.84$ & $0.84$ & $0.84$ & $0.84$\\
        $q_1$   & 0.6 & 0.6 & 0.6 & 0.6 & 0.6 & 0.6 & 0.6 & 0.6 & 0.6 & 0.6 & 0.6 & 0.6 & 0.6 & 0.6 & 0.6 & 0.6 & 0.45 & 0.45 & 0.45 & 0.45 & 0.45 & 0.45 & 0.45 & 0.45 & 0.45 & 0.45 & 0.45 & 0.45 & 0.45 & 0.45 & 0.45 & 0.45\\ 
        $q_2$  & 0.4 & 0.4 & 0.4 & 0.4 & 0.4 & 0.4 & 0.4 & 0.4 & 0.4 & 0.4 & 0.4 & 0.4 & 0.4 & 0.4 & 0.4 & 0.4 & 0.35 & 0.35 & 0.35 & 0.35 & 0.35 & 0.35 & 0.35 & 0.35 & 0.35 & 0.35 & 0.35 & 0.35 & 0.35 & 0.35 & 0.35 & 0.35\\
        \hline
        $R_{\rm p}$ & 
        $R_{\rm J}$ & $R_{\rm J}$ & $R_{\rm J}$ & $R_{\rm J}$ &
        $R_{\rm S}$ & $R_{\rm S}$ & $R_{\rm S}$ & $R_{\rm S}$ &
        $R_{\rm N}$ & $R_{\rm N}$ & $R_{\rm N}$ & $R_{\rm N}$ &
        $R_{\oplus}$ & $R_{\oplus}$ & $R_{\oplus}$ & $R_{\oplus}$ &
        $R_{\rm J}$ & $R_{\rm J}$ & $R_{\rm J}$ & $R_{\rm J}$ &
        $R_{\rm S}$ & $R_{\rm S}$ & $R_{\rm S}$ & $R_{\rm S}$ &
        $R_{\rm N}$ & $R_{\rm N}$ & $R_{\rm N}$ & $R_{\rm N}$ &
        $R_{\oplus}$ & $R_{\oplus}$ &$R_{\oplus}$ & $R_{\oplus}$\\
        $M_{\rm p}$  &
        $M_{\rm J}$ & $M_{\rm J}$ & $M_{\rm J}$ & $M_{\rm J}$ &
        $M_{\rm S}$ & $M_{\rm S}$ & $M_{\rm S}$ & $M_{\rm S}$ &
        $M_{\rm N}$ & $M_{\rm N}$ & $M_{\rm N}$ & $M_{\rm N}$ &
        $M_{\oplus}$ & $M_{\oplus}$ & $M_{\oplus}$ & $M_{\oplus}$ &
        $M_{\rm J}$ & $M_{\rm J}$ & $M_{\rm J}$ & $M_{\rm J}$ &
        $M_{\rm S}$ & $M_{\rm S}$ & $M_{\rm S}$ & $M_{\rm S}$ &
        $M_{\rm N}$ & $M_{\rm N}$ & $M_{\rm N}$ & $M_{\rm N}$ &
        $M_{\oplus}$ & $M_{\oplus}$ & $M_{\oplus}$ & $M_{\oplus}$\\
        $P_{\rm b}$ [d]& 30 & 30 & $58.31$ &$58.41$ & 30 & 30 & $58.31$ &$58.41$ & 30 & 30 & $58.31$ &$58.41$ & 30 & 30 & $58.31$ &$58.41$ & 30 & 30 & $58.31$ &$58.41$ & 30 & 30 & $58.31$ &$58.41$ & 30 & 30 & $58.31$ &$58.41$ & 30 & 30 & $58.31$ &$58.41$ \\
        $b_{\rm b}$ & 0.1 & 0.1 & 0.1 & 0.1 & 0.1 & 0.1 & 0.1 & 0.1 & 0.1 & 0.1 & 0.1 & 0.1 & 0.1 & 0.1 & 0.1 & 0.1 & 0.1 & 0.1 & 0.1 & 0.1 & 0.1 & 0.1 & 0.1 & 0.1 & 0.1 & 0.1 & 0.1 & 0.1 & 0.1 & 0.1 & 0.1 & 0.1 \\
        $\varphi_{\rm b}$  & 0 & 0 & 0 & 0 & 0 & 0 & 0 & 0  & 0 & 0 & 0 & 0 & 0 & 0 & 0 & 0  & 0 & 0 & 0 & 0 & 0 & 0 & 0 & 0  & 0 & 0 & 0 & 0 & 0 & 0 & 0 & 0 \\
        \hline
        $R_{\rm s}$  & $R_{\rm N}$  & $R_{\rm N}$  & $R_{\rm N}$  & $R_{\rm N}$ & 2~$R_{\oplus}$ & 2~$R_{\oplus}$ & 2~$R_{\oplus}$ & 2~$R_{\oplus}$ & $R_{\oplus}$ & $R_{\oplus}$ & $R_{\oplus}$ & $R_{\oplus}$ & $R_{\leftmoon}$ & $R_{\leftmoon}$ & $R_{\leftmoon}$ & $R_{\leftmoon}$ & $R_{\rm N}$  & $R_{\rm N}$  & $R_{\rm N}$  & $R_{\rm N}$ & 2~$R_{\oplus}$ & 2~$R_{\oplus}$ & 2~$R_{\oplus}$ & 2~$R_{\oplus}$ & $R_{\oplus}$ & $R_{\oplus}$ & $R_{\oplus}$ & $R_{\oplus}$ & $R_{\leftmoon}$ & $R_{\leftmoon}$ & $R_{\leftmoon}$ & $R_{\leftmoon}$\\
        $M_{\rm s}$  & $M_{\rm N}$  & $M_{\rm N}$  & $M_{\rm N}$  & $M_{\rm N}$ & 8~$M_{\oplus}$ & 8~$M_{\oplus}$ & 8~$M_{\oplus}$ & 8~$M_{\oplus}$ & $M_{\oplus}$ & $M_{\oplus}$ & $M_{\oplus}$ & $M_{\oplus}$ & $M_{\leftmoon}$ & $M_{\leftmoon}$ & $M_{\leftmoon}$ & $M_{\leftmoon}$ & $M_{\rm N}$ & $M_{\rm N}$ & $M_{\rm N}$  & $M_{\rm N}$ & 8~$M_{\oplus}$ & 8~$M_{\oplus}$ & 8~$M_{\oplus}$ & 8~$M_{\oplus}$ & $M_{\oplus}$ & $M_{\oplus}$ & $M_{\oplus}$ & $M_{\oplus}$ & $M_{\leftmoon}$ & $M_{\leftmoon}$ & $M_{\leftmoon}$ & $M_{\leftmoon}$\\
        $P_{\rm s}$  &  $P_{\rm Io}$ & $P_{\rm Eu}$ & $P_{\rm Io}$ & $P_{\rm Eu}$ & $P_{\rm Io}$ & $P_{\rm Eu}$ & $P_{\rm Io}$ & $P_{\rm Eu}$ & $P_{\rm Io}$ & $P_{\rm Eu}$ & $P_{\rm Io}$ & $P_{\rm Eu}$ & $P_{\rm Io}$ & $P_{\rm Eu}$ & $P_{\rm Io}$ & $P_{\rm Eu}$ & $P_{\rm Io}$ & $P_{\rm Eu}$ & $P_{\rm Io}$ & $P_{\rm Eu}$ & $P_{\rm Io}$ & $P_{\rm Eu}$ & $P_{\rm Io}$ & $P_{\rm Eu}$ & $P_{\rm Io}$ & $P_{\rm Eu}$ & $P_{\rm Io}$ & $P_{\rm Eu}$ & $P_{\rm Io}$ & $P_{\rm Eu}$ & $P_{\rm Io}$ & $P_{\rm Eu}$\\
        $\varphi_{\rm s}$ & 0 & 0 & 0 & 0 & 0 & 0 & 0 & 0  & 0 & 0 & 0 & 0 & 0 & 0 & 0 & 0  & 0 & 0 & 0 & 0 & 0 & 0 & 0 & 0  & 0 & 0 & 0 & 0 & 0 & 0 & 0 & 0\\
        $i_{\rm s}$ & 0.233 & 0.149 & 0.233 & 0.149 & 0.255 & 0.163 & 0.255 & 0.163 & 0.201 & 0.130 & 0.201 & 0.130 & 0.141 & 0.092 & 0.141 & 0.092 & 0.233 & 0.149 & 0.233 & 0.149 & 0.255 & 0.163 & 0.255 & 0.163 & 0.201 & 0.130 & 0.201 & 0.130 & 0.141 & 0.092 & 0.141 & 0.092 
    \end{tabular}
    \caption{Summary of our star, planet, and moon parameter grid. To preserve the orbital advancement per transit we changed the long planet orbit period from 60\,d to the given values depending on the orbital period of the moon. The inclinations ($i_{\rm s}$) tested in cases 1a - 1p are 0 and 0.2 radians. For the other cases the inclination is calculated as the smallest value to prevent occultations. The mass and radius of the Earth's moon are referred to as $M_{\leftmoon}$ and $R_{\leftmoon}$, respectively. Subscripts S and N refer to Saturn and Neptune, respectively.}
    \label{tab:params}
\end{sidewaystable}

\clearpage
\section{Plots of cases 2-9}
\label{sec:plots_plots_plots}

\begin{sidewaysfigure}
    \centering
    \includegraphics[width=.49\textwidth]{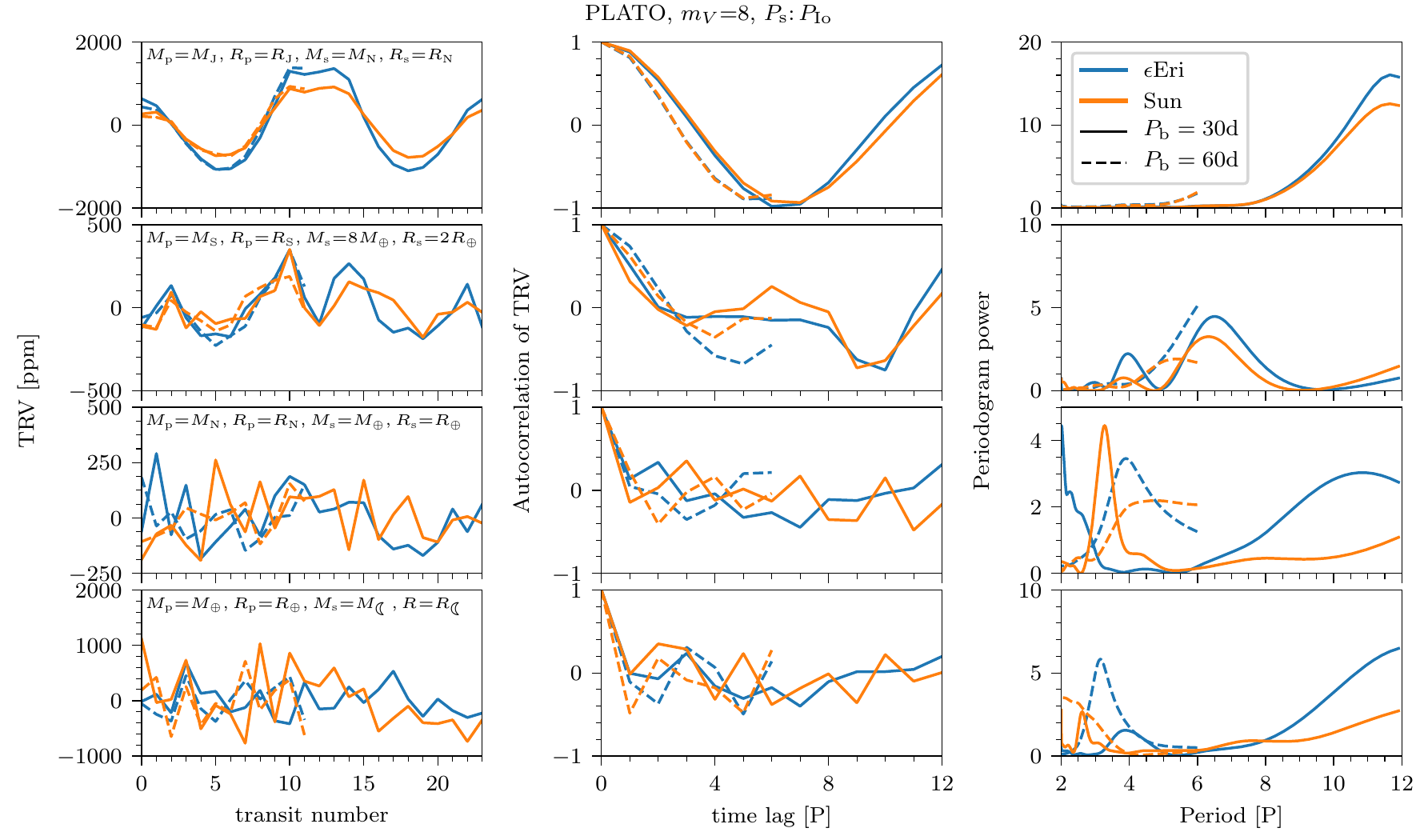}
    
    \includegraphics[width=.49\textwidth]{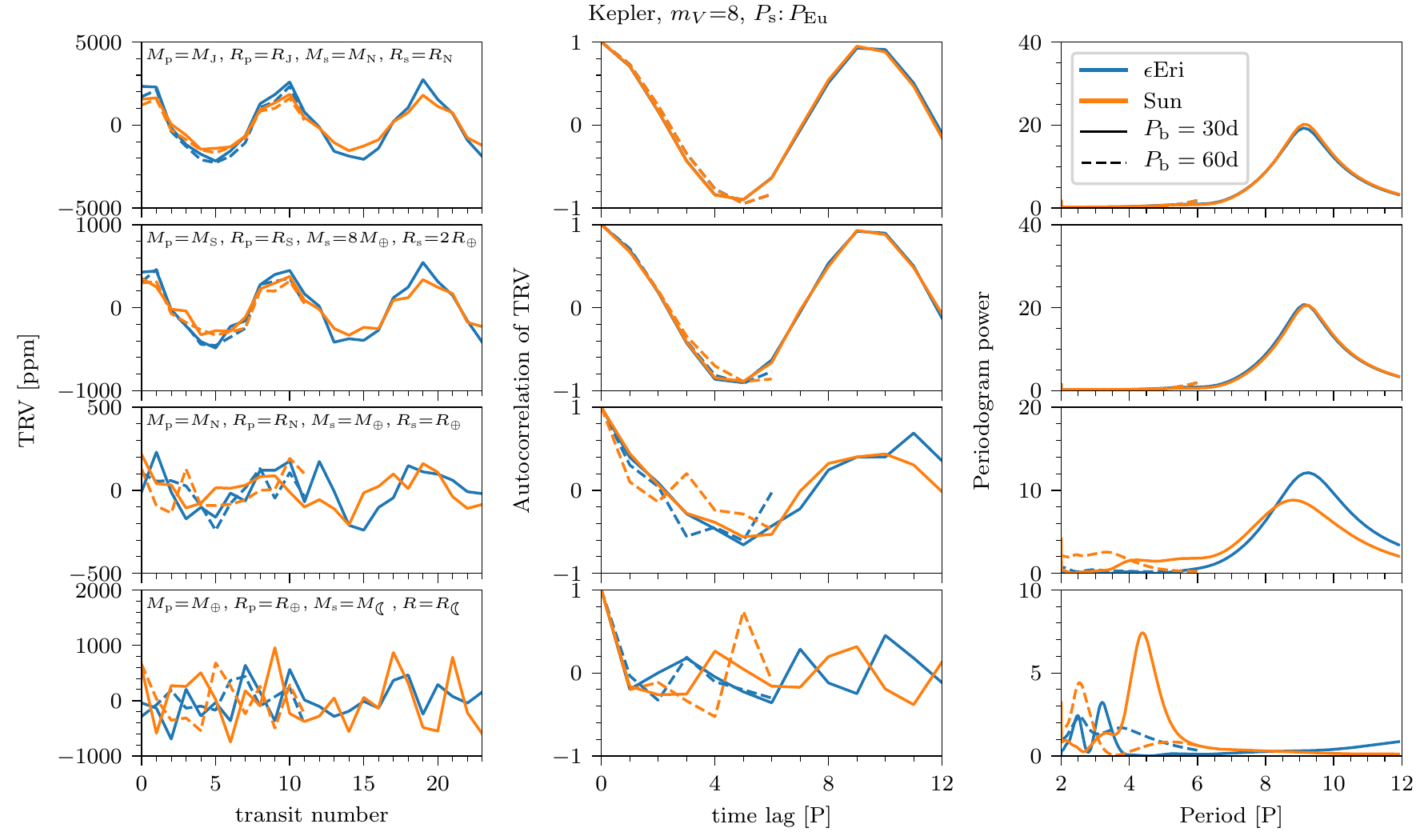}
    \includegraphics[width=.49\textwidth]{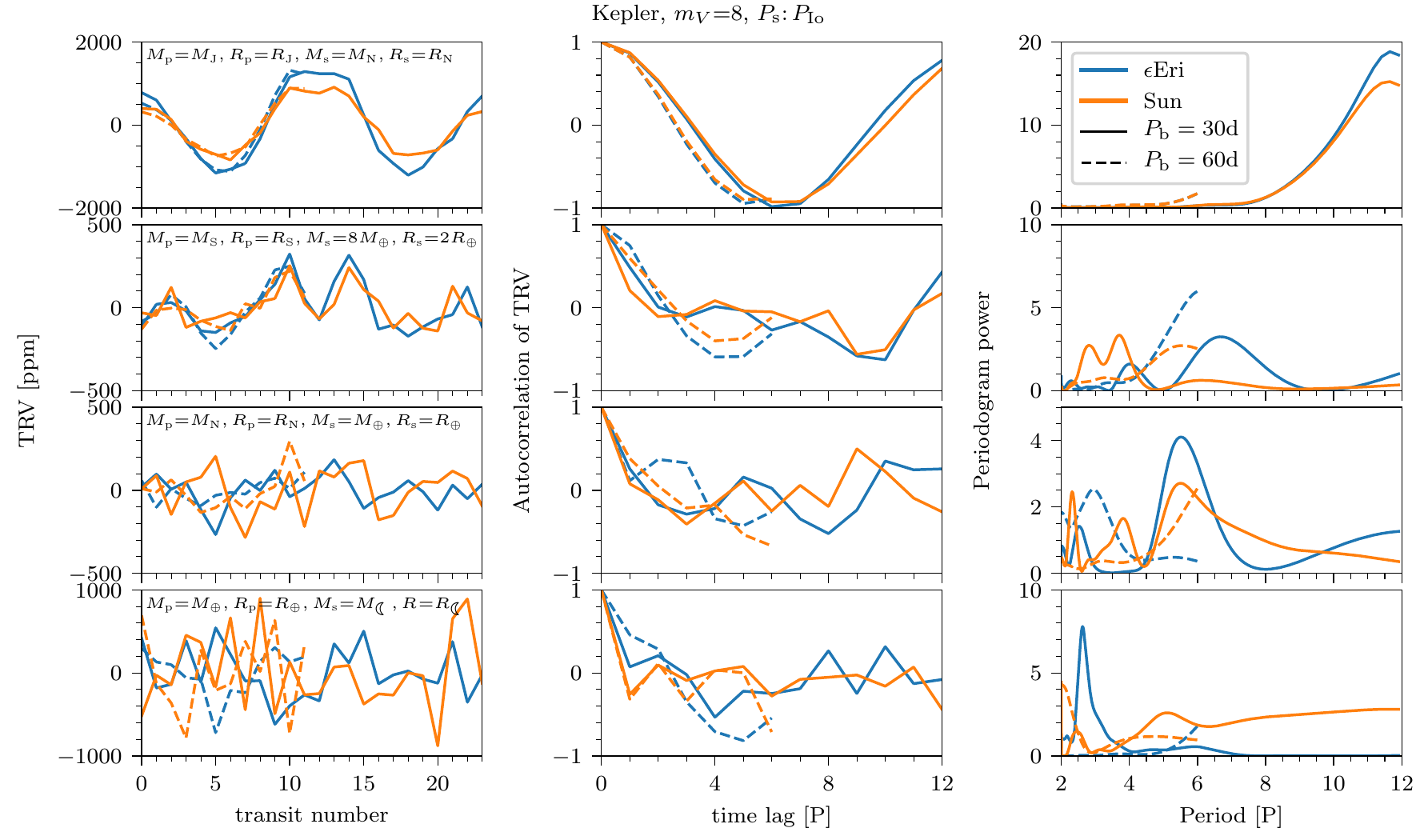}
    \caption{As Fig. \ref{fig:combined_systems_all_metrics}, but for a PLATO-like telescope and a moon period equal to Io's orbit around Jupiter (top panel) and Kepler-like telescopes and moon periods $P_{\rm Eu}$ and $P_{\rm Io}$ (bottom panels).}
    \label{fig:combined_systems_all_metrics_magn_8}
\end{sidewaysfigure}

\setlength{\rotFPtop}{0.51\textwidth}
\begin{sidewaysfigure}
    \centering
    \includegraphics[width=.49\textwidth]{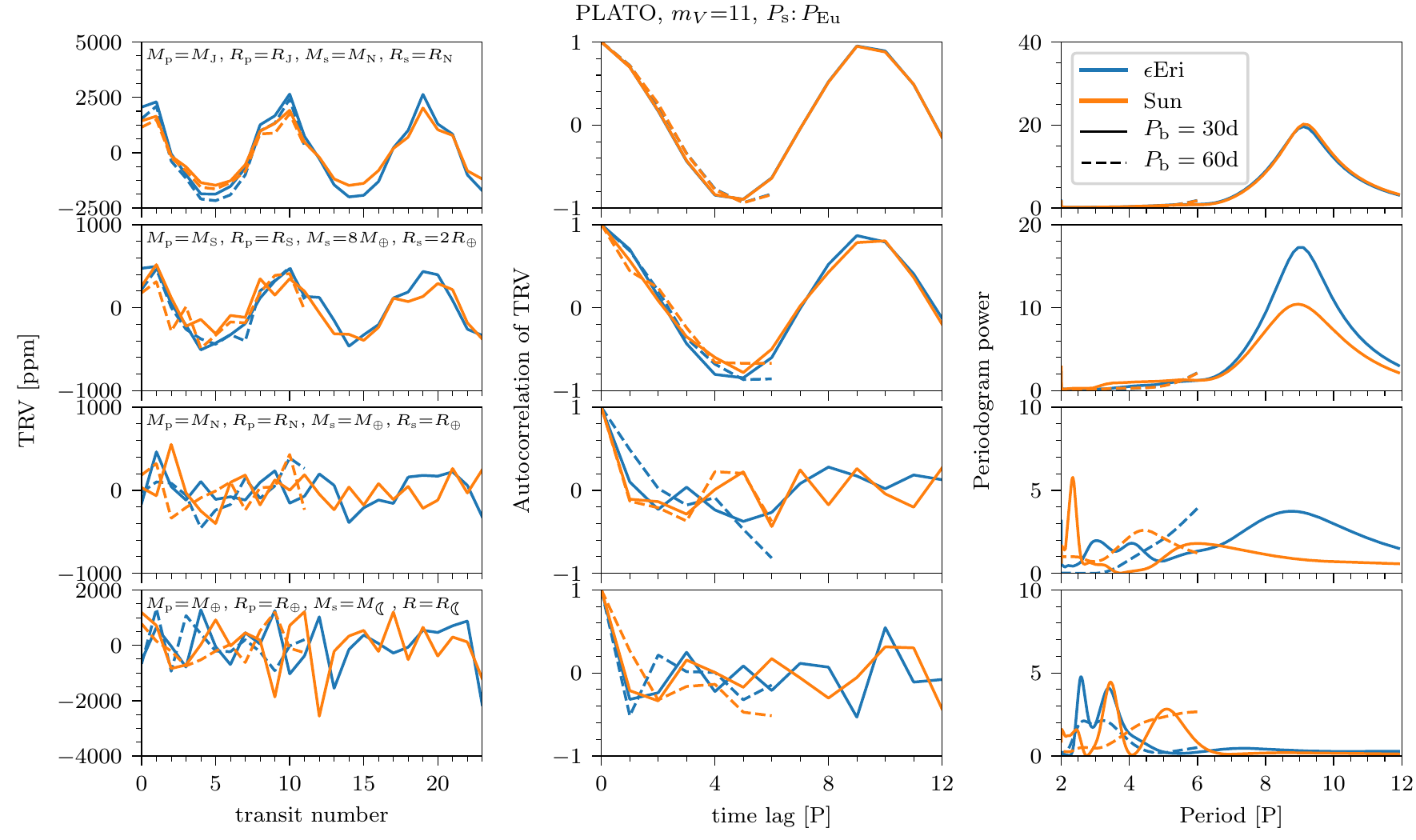}
    \includegraphics[width=.49\textwidth]{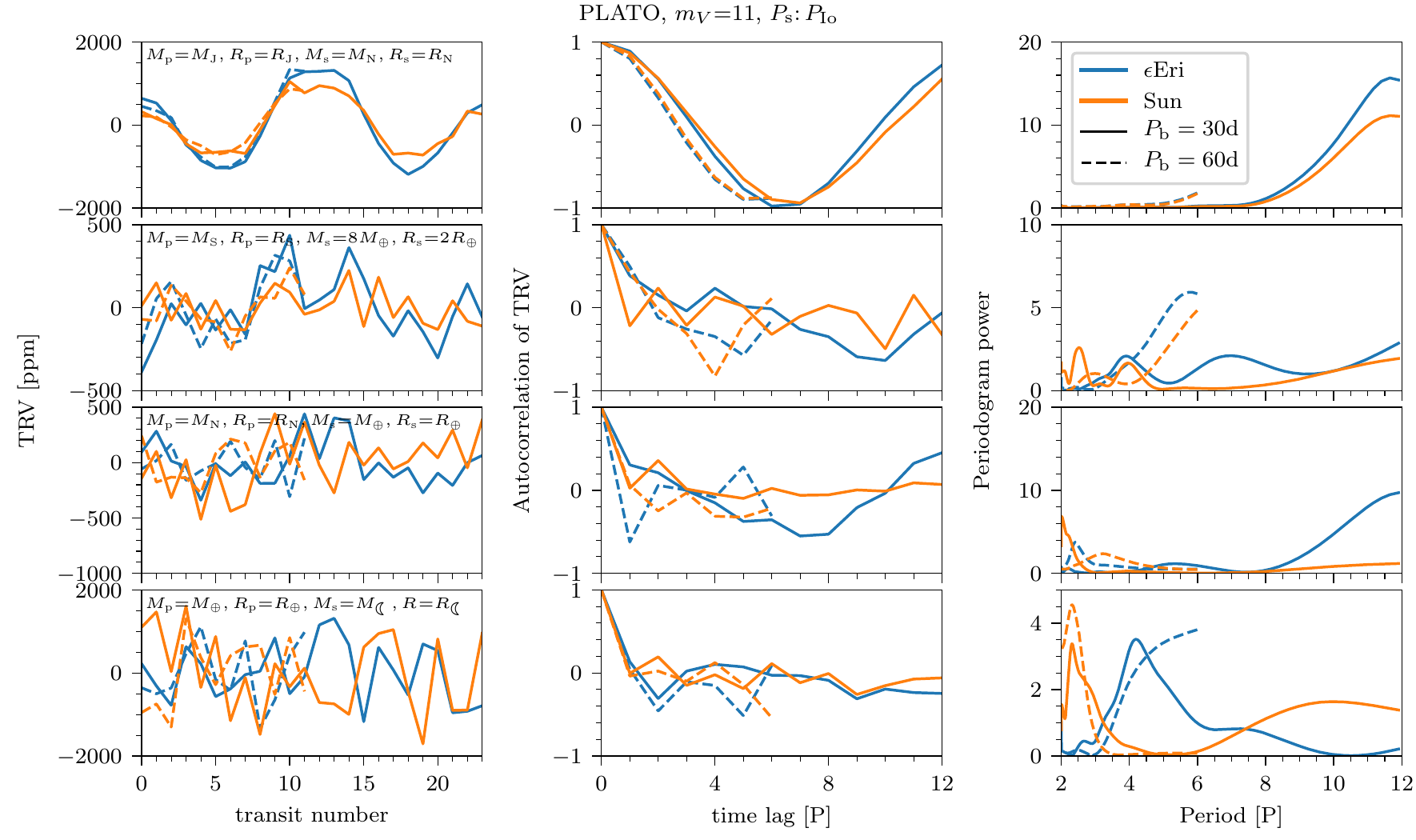}
    
    \includegraphics[width=.49\textwidth]{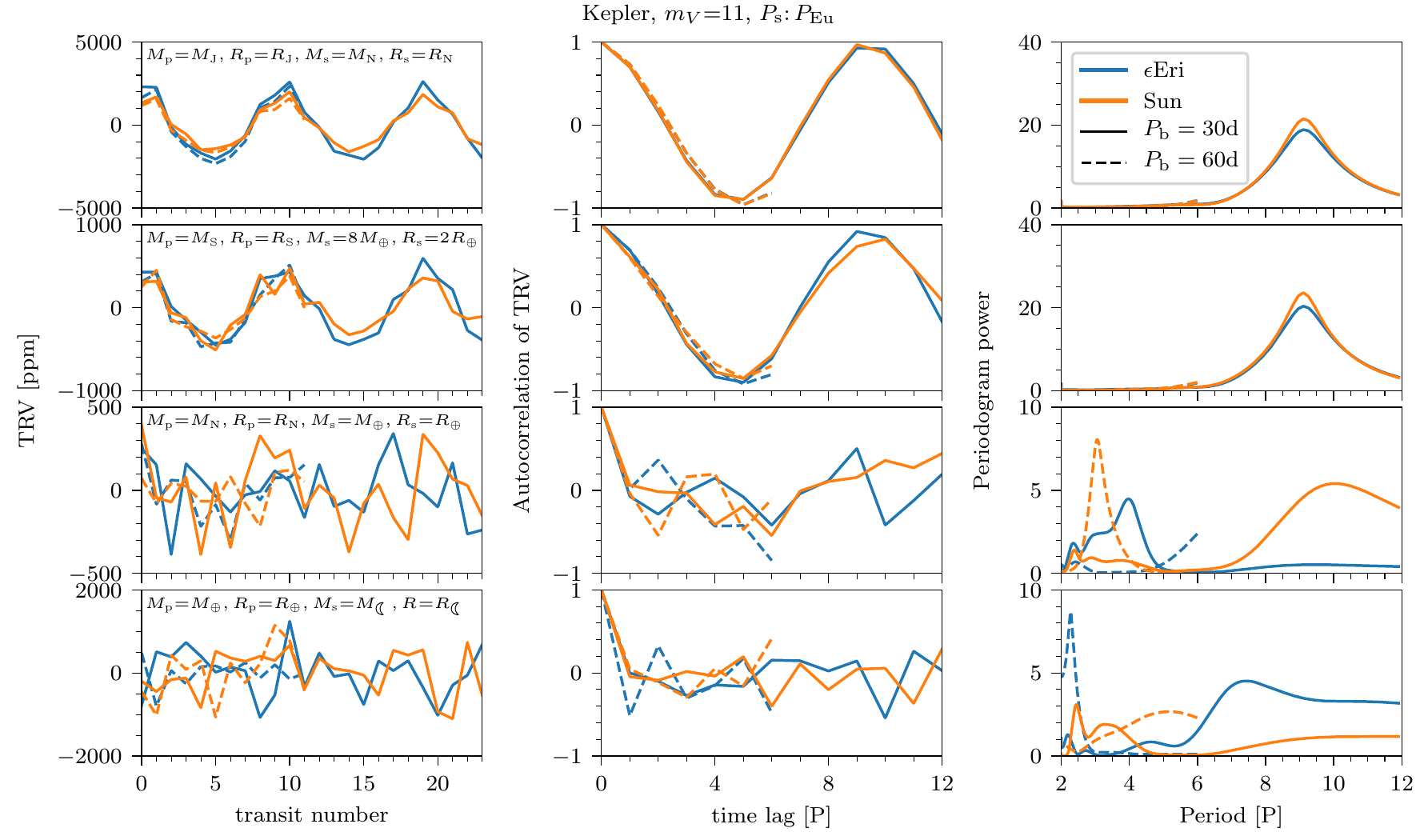}
    \includegraphics[width=.49\textwidth]{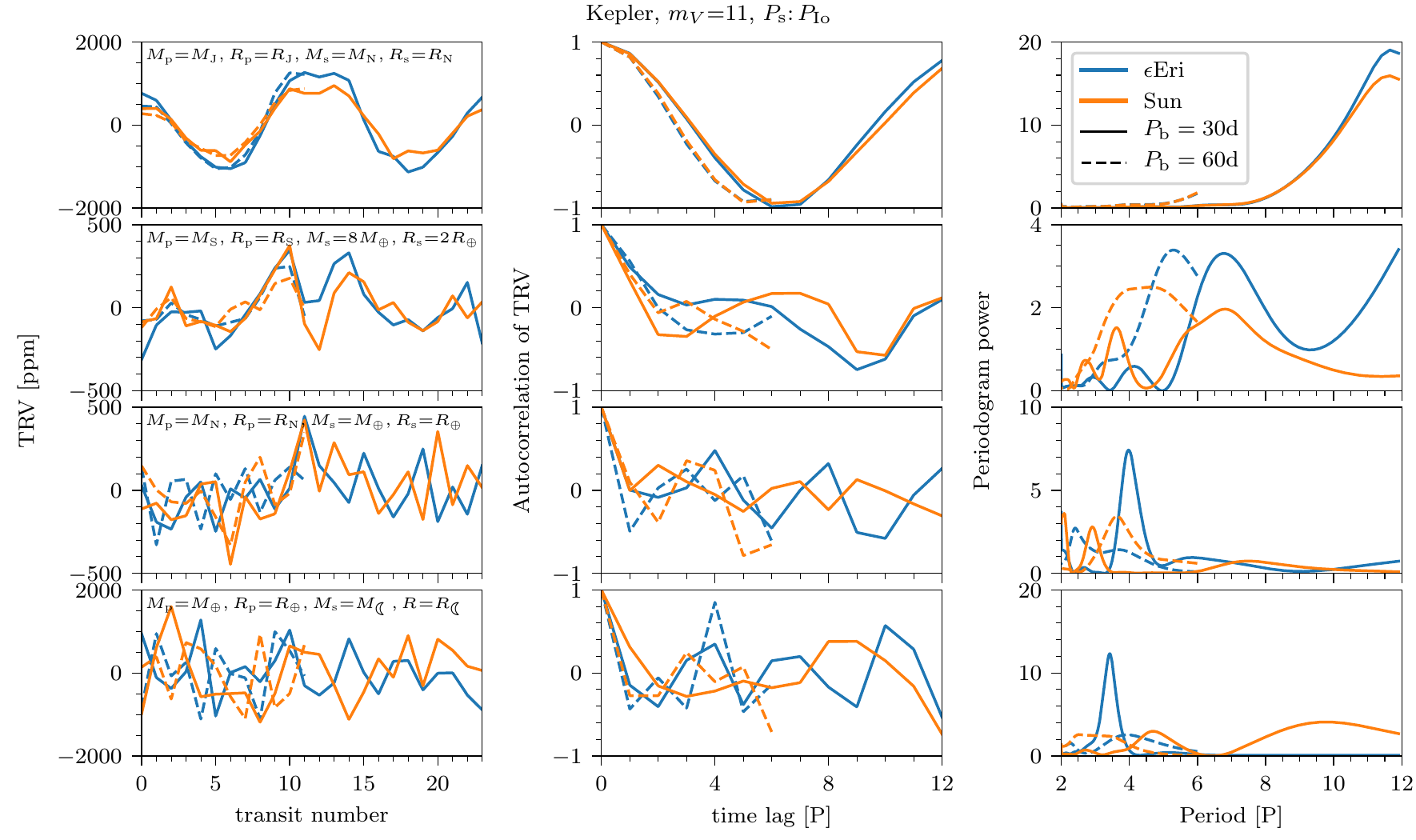}
    \caption{As Figs. \ref{fig:combined_systems_all_metrics} and \ref{fig:combined_systems_all_metrics_magn_8}, but for magnitude 11, PLATO and Kepler-like telescopes and moon periods $P_{\rm Eu}$ and $P_{\rm Io}$.}
    \label{fig:combined_systems_all_metrics_magn_11}
\end{sidewaysfigure}

\clearpage
\section{Aliasing of an observed TRV}
\label{sec:aliasing}

In our model of a star--planet--moon system, the phase of the moon orbit around the planet advances by a constant fraction $0~\leq~r~\leq~1$ between successive transits although the moon orbits the planet many times between the transits. This advancement can be calculated from the orbital period of the planet-moon barycenter around the star ($P_{\rm b}$) and the orbital period of the satellite around the local planet-moon barycenter ($P_{\rm s}$) as

\begin{align}
    \frac{P_{\rm b}}{P_{\rm s}}=n+r \ ,
\end{align}

\noindent
where $n$ is the number of completed orbits between successive transits and $r$ is the remainder \citep{2016A&A...591A..67H}. In a similar way, the remainder of the TRV signal $r_{\rm TRV}$ can be calculated. But note that the period of the TRV effect ($P_{\rm TRV}$) is actually half of the orbital period of the moon, $P_{\rm TRV}~=~P_{\rm s}/2$, as can be seen in Fig.~\ref{fig:TTV_TDV_TRV_photometric_barycentric_moon_phase} for example.\footnote{This is due to the projection of the three-dimensional moon orbit onto the two-dimensional celestial plane \citep[see Fig.~1 in][]{2014ApJ...796L...1H}.} The frequency of the TRV signal then is

\begin{align}
f_{\rm TRV} = \frac{1}{P_{\rm TRV}} = \frac{2}{P_{\rm s}} \ .
\end{align}

\noindent
The TRV signal can only be sampled during transits, a constraint that establishes a sampling frequency $f_{\rm sam}=1/P_{\rm b}$ and a Nyquist frequency $f_{\rm Nyq}=f_{\rm sam}/2$ .

If $\left|f_{\rm TRV}\right|$ is larger than $f_{\rm Nyq}$ then the signal is aliased and a full integer multiple ($N$) of $f_{\rm sam}$ is subtracted from $f_{\rm TRV}$ so that the resulting alias frequency $f_{\rm TRV, alias}$ falls between $-f_{\rm Nyq}$ and $+f_{\rm Nyq}$,

\begin{align}
\label{eq:alias}
f_{\rm TRV, alias}=f_{\rm TRV}-Nf_{\rm sam} \ , \ \ {\rm with}~N={\rm round}\left(\frac{f_{\rm TRV}}{f_{\rm sam}}\right) \ .
\end{align}

\noindent
The resulting period of the alias TRV signal then comes out as

\begin{align} \label{eq:P_TRV} \nonumber
P_{\rm TRV,alias} &= \frac{1}{f_{\rm TRV,alias}} = \frac{1}{f_{\rm TRV} - N f_{\rm sam}} = \frac{1}{\frac{\displaystyle 2}{\displaystyle P_{\rm s}} - \frac{\displaystyle N}{\displaystyle P_{\rm b}}} \\
                  &= \frac{1}{\frac{\displaystyle 2}{\displaystyle P_{\rm s}} - \frac{\displaystyle N}{\displaystyle P_{\rm b}}} \ .
\end{align}

As an example, let us consider the case of a Jupiter-sized planet in a $P_{\rm b}=30$\,d orbit around a sun-like star and an Earth-sized moon in an orbit with a Europa-like orbital period of $P_{\rm s}~=~3.5512$\,d as shown in Fig.~\ref{fig:shifted_noVIRGO_jupiter_earth}. We find $N=17$ and then use Eq.~\eqref{eq:P_TRV} to calculate an alias period of the TRV signal as 287.6\,d, corresponding to $9.59\,P_{\rm b}$. This is where the peak in the TRV periodogram presented in Fig.~\ref{fig:shifted_noVIRGO_jupiter_earth} occurs.

These calculations assume that the TRV signal is composed of a single sine wave. If the TRV signal is not a sine but composed of multiple sine waves (in the sense of a Fourier transformation), then each frequency of the signal is folded individually, which might lead to alias signals that have a very different form in Fourier space than the original signal.

\end{document}